\documentclass[aps,prb,twocolumn,amsmath,amssymb,showpacs,superscriptaddress,reprint, floatfix]{revtex4}
\usepackage{times}
\usepackage{graphicx}
\usepackage{color}

\begin{document}

\title{Observation of conduction electron spin resonance in boron doped diamond}

\author{P\'{e}ter~Szirmai}
\affiliation{Laboratory of Physics of Complex Matter, \'{E}cole Polytechnique F\'{e}d\'{e}rale de Lausanne, CH-1015 Lausanne, Switzerland}
\affiliation{Department of Physics, Budapest University of Technology and Economics, P.O.\ Box 91, Budapest, H-1521, Hungary}

\author{G\'{a}bor~F\'{a}bi\'{a}n}
\altaffiliation[Present address: ]{Department of Physics, University of Basel, Klingelbergstrasse 82, CH-4056 Basel, Switzerland}
\affiliation{Department of Physics, Budapest University of Technology and Economics, P.O.\ Box 91, Budapest, H-1521, Hungary}

\author{J\'{a}nos~Koltai}
\affiliation{Department of Biological Physics, E\"{o}tv\"{o}s
University, P\'{a}zm\'{a}ny P\'{e}ter s\'{e}t\'{a}ny 1/A, H-1117
Budapest, Hungary}

\author{B\'{a}lint~N\'{a}fr\'{a}di}
\affiliation{Laboratory of Physics of Complex Matter, \'{E}cole Polytechnique F\'{e}d\'{e}rale de Lausanne, CH-1015 Lausanne, Switzerland}

\author{L\'aszl\'o~Forr\'o}
\affiliation{Laboratory of Physics of Complex Matter, \'{E}cole Polytechnique F\'{e}d\'{e}rale de Lausanne, CH-1015 Lausanne, Switzerland}

\author{Thomas~Pichler}
\affiliation{Faculty of Physics, University of Vienna, Strudlhofgasse 4., Vienna, A-1090, Austria}

\author{Oliver~A.~Williams}
\affiliation{School of Physics and Astronomy, Cardiff University, Cardiff CF24 3AA, UK}

\author{Soumen~Mandal}
\affiliation{Institut N\'{e}el- CNRS and Universit\'{e} Joseph Fourier, 38042 Grenoble, France}

\author{Christopher~B\"{a}uerle}
\affiliation{Institut N\'{e}el- CNRS and Universit\'{e} Joseph Fourier, 38042 Grenoble, France}

\author{Ferenc~Simon}
\email[Corresponding author: ]{ferenc.simon@univie.ac.at}
\affiliation{Department of Physics, Budapest University of Technology and Economics, P.O.\ Box 91, Budapest, H-1521, Hungary}
\affiliation{Condensed Matter Physics Research Group of the Hungarian Academy of Sciences, P.O.\ Box 91, H-1521 Budapest, Hungary}

\date{\today}
\begin{abstract}
We observe the electron spin resonance of conduction electrons in boron doped (6400~ppm) superconducting diamond ($T_{\text{c}} =3.8~\text{K}$). We clearly identify the benchmarks of conduction electron spin resonance (CESR): the nearly temperature independent ESR signal intensity and its magnitude which is in good agreement with that expected from the density of states through the Pauli spin-susceptibility. The temperature dependent CESR linewidth weakly increases with increasing temperature which can be understood in the framework of the Elliott-Yafet theory of spin-relaxation. An anomalous and yet unexplained relation is observed between the $g$-factor, CESR linewidth, and the resistivity using the empirical Elliott-Yafet relation.\end{abstract}

\maketitle

\section{Introduction}

Information storage and processing using electron spins, referred to as spintronics,~\cite{FabianRMP,WuReview} is an ambitious proposition to provide a technological leap in information sciences.
The spin relaxation time in metals and semiconductors, $\tau_{\text{s}}$, is the central parameter which characterizes their utility for spintronic applications.~\cite{FabianRMP} There are two viable routes to determine $\tau_{\text{s}}$ in these materials: transport and spectroscopy based. Transport studies usually detect the decay length of an injected non-equilibrium magnetization of a spin ensemble in a non-local resistivity measurement.~\cite{TombrosNAT2007,GuntherodtBilayer,KawakamiBilayer} The characteristic decay or spin-diffusion length, $\delta_{\text{spin}}$, contains $\tau_{\text{s}}$ through: $\delta_{\text{spin}}=v_{\text{F}}\sqrt{\tau \cdot \tau_{\text{s}}/d}$, where $d$ is the dimensionality of the material, $v_{\text{F}}$ is the Fermi velocity and $\tau$ is the momentum scattering time. Electron spin resonance (ESR) experiments in metals or semiconductors~\cite{KipKittelPR1952} are also capable of determining $\tau_{\text{s}}$ from the homogeneous ESR linewidth, $\Delta B$, through $\tau_{\text{s}}=1/\gamma \Delta B$, where $\gamma/2\pi=28.0~\text{GHz/T}$ is the electron gyromagnetic ratio. A spectroscopy related but transport based method is the so-called Hanle spin-precession experiment which also yields $\tau_{\text{s}}$ values.~\cite{FabianRMP}\\
In metals with inversion symmetry the so-called Elliott-Yafet theory describes the spin-relaxation properties. It predicts that the spin-relaxation time is proportional to the momentum relaxation time ($\tau_{\text{s}}\propto \tau$) and thus the ESR linewidth is proportional to the resistivity ($\Delta B\propto \varrho$).

Heavily boron doped diamond (BDD) is an example of Mott's metal~\cite{Mott1968} above the threshold boron concentration of $n_c\approx 4\text{-}5\cdot 10^{20}~\text{cm}^{-3}$.~\cite{BDD_KleinPRB, BDD_Bustarret2008, Carrier} The discovery of superconductivity in BDD~\cite{BDD_Nature2004} attracted significant interest and it has been proven that superconductivity in BDD is an intrinsic property which arises from the lightly hole doped diamond bands and not due to the acceptor bands/levels.~\cite{BDD_Nature2005}\\
Diamond possesses a number of unique properties (such as e.g.\ the well-known hardness, large tensile strength, and thermal conductivity) which may lead to a unique class of fully diamond based integrated (or even spintronic) devices.~\cite{BDD_Mandal} Clearly, knowledge of the spin-relaxation time in this new metal is a prerequisite for such applications. In addition, theory of spin-relaxation in metals and semiconductors is an ever developing field which progresses by testing the basic theories against new materials.

Herein, we study the electron spin resonance in superconducting BDD ($T_{\text{c}}=3.8~\text{K}$) in the $5\text{-}300~\text{K}$ temperature range, i.e.\ in the normal state. We observe three ESR signals with a temperature dependence which is characteristic for localized paramagnetic centers. In addition, an ESR signal, which is assigned to conduction electrons, is observed. The identification is made by examining the signal intensity and its temperature dependence, which cannot be explained by localized spins. The calibrated signal intensity yields the Pauli spin-susceptibility whose experimental value agrees with that obtained from the band structure-based density of states data. The CESR linewidth increases with increasing temperature which is characteristic for the Elliott-Yafet spin-relaxation mechanism.~\cite{Elliott,YafetPL1983} Measurement of the $g$-factor and the linewidth allows to place BDD on the empirical Beuneu-Monod plot ,~\cite{BeuneuMonodPRB1978} which summarizes the spin-relaxation properties of elemental metals.

\section{Experimental}

We performed the ESR experiments on powders of silicon-wafer free boron doped diamond samples. To prepare the samples, silicon $(111)$ wafers were cleaned by standard RCA SC1 processes. Diamond nucleation was initiated by immersion of clean wafers in aqueous colloids of hydrogenated nanodiamond particles in an ultrasonic bath. This process is known to produce nucleation densities in excess of $10^{11}~\text{cm}^{-2}$.~\cite{BDD_Hees} Diamond growth for 20 hours using Microwave Plasma Enhanced Chemical Vapour Deposition with 4\% CH$_4$ diluted in H$_2$ with 6400~ppm of trimethylboron,~\cite{BDD_Gajewski} microwave power of 3~kW and substrate temperature of $800~^{\circ}$C yields films of approximately $6~\mu\text{m}$ thickness.\\

Characterization by transport experiments on BDD films found a $T_{\text{c}}=3.8~\text{K}$.~\cite{SupMat} $T_{\text{c}} \approx 4~\text{K}$ usually corresponds to a boron concentration of $n \approx 10^{21}~{\text {cm}}^{-3}$ (or $\sim 6000~\text{ppm}$) according to the calibration established for samples prepared with chemical vapour deposition.~\cite{BDD_NMR_conc, BDD_CVD_Diamond2007} The Si substrate was removed using a mixture of HF and H$_2$SO$_4$. The former oxidizes Si, the latter removes SiO$_2$.

Experiments at 9 GHz (0.33 T) were carried out on finely ground BDD samples using a Bruker Elexsys E500 ESR spectrometer in the $5\text{-}300~\text{K}$ temperature range. Care was taken to employ large magnetic modulation to enhance the broad resonance signal due to the itinerant electrons and also to eliminate any spurious background signals from the cavity or the cryostat. We used Mn:MgO with a known (1.5~ppm) Mn$^{2+}$ concentration as a $g$-factor and intensity standard and we also compared the spin-susceptibility, $\chi_{\text{s}}$, against KC$_{60}$ which has a known $\chi_{\text{s}}=8\cdot 10^{-4}~\text{emu/mol}$.~\cite{KC60_Dimer_PRB1995} The ESR spectra are deconvoluted into a sum of derivative Lorentzian curves, which are an admixture of dispersion and absorption lineshapes.

\section{Results and discussion}

\begin{figure}[tb]%
\includegraphics*[width=\linewidth]{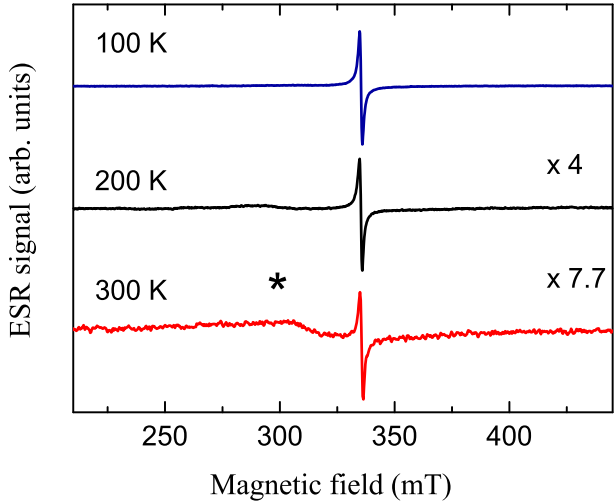}
\caption{ESR spectra of BDD at different temperatures. Note the broad ESR line (denoted by asterisk) which is assigned to the conduction electrons. The sharp ESR line originates from localized defect spins.}
\label{ESRspectra}
\end{figure}

\begin{figure}[tb]%
\includegraphics*[width=\linewidth]{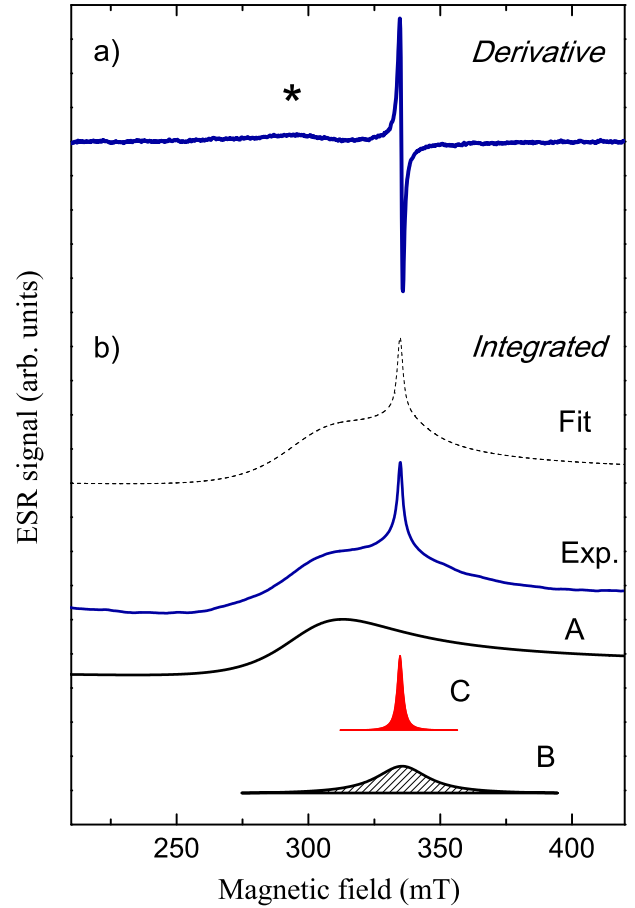}
\caption{ESR spectrum of BDD at 175 K; \textit{a)} raw, derivative data, \textit{b)} integrated ESR signal. A fit (Fit) with three components (\textit{A}, \textit{B}, and \textit{C}) simulates well the experiment (Exp.). Note the two narrow signals (\textit{B} and \textit{C}), which originate from defects and the broader component (\textit{A}) coming from the conduction electrons, which displays a Dysonian (asymmetric) lineshape.}
\label{ESRspectrum175K}
\end{figure}

\begin{figure}[tb]%
\includegraphics*[width=\linewidth]{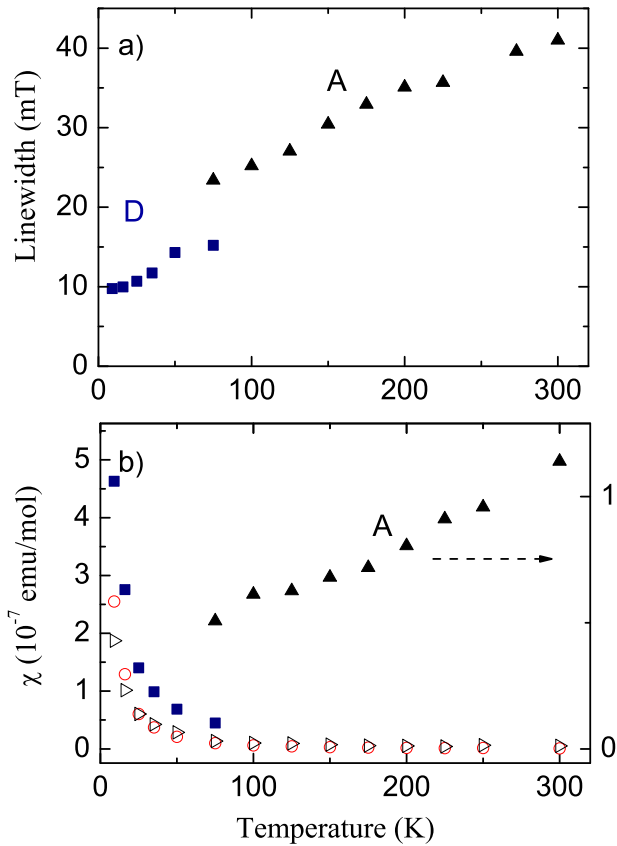}
\caption{\textit{a)} ESR linewidth of the \textit{A} and \textit{D} ESR lines as a function of temperature. \textit{b)} Spin-susceptibility as obtained from the ESR signal intensity (\textit{A}: $\blacktriangle$, \textit{B}: {$\triangleright$}, \textit{C}: \textcolor{red}{$\circ$}, and \textit{D}: \textcolor{blue}{$\blacksquare$}) as a function of temperature. The result for \textit{A} is magnified for better visibility.}
\label{ESR_linewidth_int}
\end{figure}

The identification of an ESR signal originating from the itinerant electrons in a metal relies on the following benchmarks~\cite{SzirmaiPSSB2011} in the order of importance i) the value of the measured spin-susceptibility should match the Pauli spin-susceptibility, which is related to the DOS, ii) the temperature dependence of the signal intensity should be characteristically different from the Curie, i.e. $1/T$, dependence, iii) for a metal with inversion symmetry, the linewidth should increase with increasing temperature, which is the so-called Elliott-Yafet relaxation mechanism,~\cite{Elliott,YafetPL1983} iv) the $g$-factor shift, $\Delta g=g-g_0$, and the ESR line-width should obey the so-called Elliott-Yafet relation.~\cite{Elliott,YafetPL1983}

Fig.~\ref{ESRspectra}.\ illustrates the evolution of the ESR spectrum of BDD from 100~K to 300~K. A narrow line with resonance field, $B_0\approx 335~\text{mT}$ decreases with increasing temperature, whereas the intensity of the broader line with $B_0\approx 310~\text{mT}$ does not change significantly.

In Fig.~\ref{ESRspectrum175K}a., we show the ESR spectrum of BDD at 175~K in more detail. In Fig.~\ref{ESRspectrum175K}b., the integrated spectrum is deconvoluted into three curves. In the following, we refer to these as \textit{A} ($g_A=2.16(3)$), \textit{B} ($g_B=2.003(1)$), and \textit{C} ($g_C=2.003(1)$). The \textit{B} and \textit{C} signals are assigned to bulk defects probably accompanying hydrogen vacancy complexes.~\cite{BDD_Zhou, BDD_Mizuochi3}

The \textit{A} signal dominates the integrated spectrum at 175~K due to its large linewidth. It is known that broader lines are suppressed compared to the narrower ones in the ESR technique (which uses derivative signals) as the signal amplitude drops with $1/\left(\Delta B\right)^2$. Thus the integration visually enhances the broader components.~\cite{JanossyPRL1993} The signal \textit{A} is a strongly asymmetric Lorentzian (known as Dysonian), with an equal mixture of dispersion and absorption components. This ESR line-shape is encountered in metals~\cite{FeherKip} when the itinerant electrons relax their spin state while diffusing through the penetration depth.~\cite{SupMat} The other, impurity-related signals do not show a pronounced asymmetry except at lower temperatures (below 35~K) which suggests that the corresponding spins are concentrated close to the grain surfaces.\\

Below 75~K, we identify a further ESR signal (signal \textit{D}) with a $g_D=2.016(1)$ $g$-factor.~\cite{SupMat} Its origin is discussed below. Deconvolution of the ESR spectra into several components varies in the different temperature ranges. The \textit{D} signal can be followed up to 75~K, however, starting from 75~K the fit converges to the signal \textit{A}. At 75~K the two signals can be fitted independently. Fig.~\ref{ESR_linewidth_int}a.\ depicts the ESR linewidth of both \textit{A} and \textit{D} signals as a function of temperature. The linewidth of \textit{A} is 10~mT larger than that of \textit{D} at $T=75~\text{K}$. Hence, we find that the \textit{A} and \textit{D} signals have different origins. Below 75~K the \textit{A} signal cannot be resolved, whose origin is unexplained. We speculate that this effect is caused by weak-localization (WL), which may either lead to a sudden line broadening or a loss of spin-susceptibility. It is known that WL becomes significant in BDD below around 100-150~K,~\cite{BDD_KleinPRB} which supports that the change of the ESR signal of itinerant electrons and WL may be related. The linewidth of the \textit{A} signal weakly increases with temperature and it has a sizeable residual value. These observations are in agreement with the Elliott-Yafet theory of spin-relaxation. In addition, the $\Delta g$ is positive for BDD, which is compatible with hole nature of charge carriers in BDD. \\

In Fig.~\ref{ESR_linewidth_int}b., the spin-susceptibility of the four ESR signals are shown. \textit{B}, \textit{C}, and \textit{D} exhibit a Curie ($\chi_{\text{s}} \propto T^{-1}$) temperature dependence which is characteristic for localized, paramagnetic centers. Their total spin concentration corresponds to about 2.2~ppm per carbon atom in our sample,~\cite{SupMat} their presence is therefore not expected to substantially modify the intrinsic properties of BDD similar to the hydrogen vacancy complex.~\cite{BDD_Mizuochi2} The ESR intensity of \textit{A} increases by a factor two in the temperature range of 75~K to 300~K. This increase rules out that this signal would originate from localized spins. Instead, its most probable origin is the itinerant conduction electrons in BDD. A similar increase of the CESR signal intensity with a factor of 2-3 was observed in granular MgB$_2$ samples in the 40-300~K temperature range.~\cite{SimonPRL2001,SimonPRB2005b} Therein, this effect was explained by the limited microwave penetration in the metallic grains: on increasing temperature the microwave penetration depth increases due to the increasing resistivity thus resulting in an increasing CESR signal.

Given this slight uncertainty due to the limited microwave penetration, we regard the room temperature CESR signal intensity and associate it with the Pauli spin-susceptibility of the itinerant electrons. Calibration of the \textit{A} signal intensity yields $D(E_{{\text{F}}}) =4(1)\cdot 10^{-3}~\text{states/}(\text{eV}\cdot\text{C-atom})$ for the density of states of BDD.
This value corresponds to a Pauli susceptibility $\chi_{\text{s}}(\textrm{Pauli})=\frac{g^2}{4} \mu_{\textrm{B}}^2D(E_{\textrm F})\cdot N_{\textrm{A}}$ of $ 1.3(3)\cdot 10^{-7}~\text{emu}/\text{mol}=1.1(3)\cdot 10^{-8}~\text{emu}/\text{g}$ (here $N_{\text{A}}$ is the Avogrado number and $ \mu_{\text{B}}$ is the Bohr magneton).\\
This value is about two orders of magnitude lower spin-susceptibility as compared to other metallic carbon phases such as e.g.\ K$_3$C$_{60}$ ($\chi_{\text{s}}\approx 10^{-6}~\text{emu}/\text{g}$, Ref.~\onlinecite{GunnRMP}) or the KC$_8$ alkali intercalated graphite $\chi_{\text{s}}\approx 6.4\cdot 10^{-7}~\text{emu}/\text{g}$ (Ref.~\onlinecite{DresselhausAP2002}) due to the small carrier density.


\begin{figure}[tb]
\includegraphics*[width=0.9\linewidth,height=0.73\linewidth]{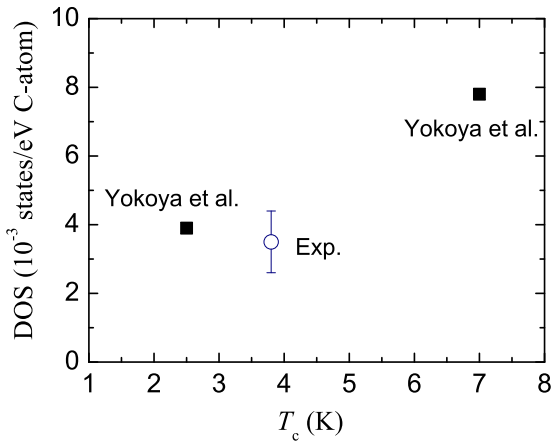}
\caption{Density of electronic states in BDD versus $T_{\text{c}}$. Experimental DOS of the present work (\textcolor{blue}{$\circ$}) is shown together with DOS calculated from the ARPES measurements in Ref.~\onlinecite{BDD_Nature2005} ($\blacksquare$). Error bar in our experiment is a conservative estimate and considers the uncertainty due to the limited microwave penetration depth.}
\label{DOS}
\end{figure}

\begin{figure}[tb]%
\includegraphics*[width=\linewidth]{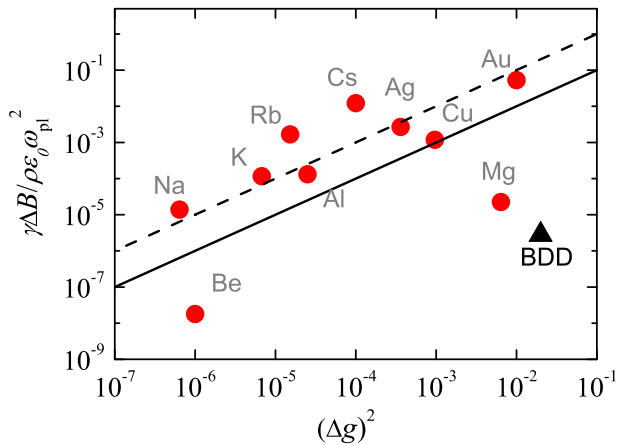}
\caption{ $\gamma \Delta B/\varrho \varepsilon_0 \omega_{\text{pl}}^2$ as a function of $\Delta g^2$ (corrected Beuneu-Monod plot~\cite{Fabian2012})for elemental metals~\cite{BeuneuMonodPRB1978} (\textcolor{red}{$\bullet$}) and BDD ($\blacktriangle$). The resistivity data for BDD is taken from Ref.~\onlinecite{BDD_Mares}. Solid and dashed lines correspond to $\alpha=1$ and 10, respectively. We use a plasma frequency of $\omega_{\text{pl}}=0.8~\text{eV}$ after Ref.~\onlinecite{BDD_Ortolani2006}.}
\label{BMplot}
\end{figure}

As mentioned above, whether the absolute magnitude of the Pauli spin-susceptibility and the corresponding DOS matches the theoretical estimates and other experimental results is an important benchmark to identify a CESR signal. We compare the DOS determined herein with angle-resolved photoemission spectroscopy (ARPES) based DOS data and with theoretical estimates in Fermi-gas and first principles based models. The present DOS value and those based on the ARPES studies are shown in Fig.~\ref{DOS}. as a function of $T_{\text{c}}$. The ARPES based DOS data is obtained directly from the measured band-structure.
 A free-carrier concentration of $n=1.1\cdot 10^{21}~{\text{cm}}^{-3}$ which corresponds to 6400~ppm boron doping~\cite{SupMat, Carrier} was used for the Fermi-gas and first principles based DOS calculations, which gave $D(E_{{\text{F}}})=2.4\cdot 10^{-2}$ and $D(E_{{\text{F}}})=4\cdot 10^{-2}$ in units of $\text{states}/(\text{eV}\cdot\text{C-atom})$, respectively.~\cite{SupMat}

 Clearly, the ARPES based DOS is in good agreement with the present data, whereas the theoretical estimates significantly differ. We note that the effective carrier concentration in BDD is lowered by the presence of boron dimers,~\cite{BDD_Bourgeois} which may explain this difference. The presence of boron dimers justifies the use of the truly empirical DOS versus $T_{\text{c}}$ comparison.

In the following, we discuss the validity of the Elliott-Yafet relation, $\gamma \Delta B=\alpha (\Delta g)^2 \varrho \epsilon_0 \omega_{\text{pl}}^2$ ($\varrho$ is the resistivity, $\epsilon_0$ is the vacuum permittivity, and $\omega_{\text{pl}}$ is the plasma frequency), in BDD. The Elliott-Yafet relation combines three independent empirical parameters, $\Delta B$, $\Delta g$, and $\varrho$, i.e. it is a benchmark of spin-relaxation experiments in novel metals.~\cite{FabianRMP} Beuneu and Monod~\cite{BeuneuMonodPRB1978, MonodBeuneuPRB1979} verified its validity for elemental metals and established a linear scaling, i.e. the empirical constant being $\alpha \approx 1..10$.
In Fig.~\ref{BMplot}., we show the Beuneu-Monod plot together with the present results for BDD. Clearly, BDD lies out of the general trend observed for most metals. We note that an overestimate of the resistivity may contribute to this effect as the granularity of BDD samples hinders measurement of the intrinsic $\varrho$.~\cite{BDD_Zhang}

It is known that the linear scaling of the Elliott-Yafet relation occurs mostly for monovalent materials and notable exceptions are Be and Mg for which the so-called "hot-spot" model was invoked to explain the data.~\cite{FabianPRL1998} The hot-spot model recognizes that spin-relaxation is enhanced for particular points of the Fermi-surface; given that the spin life-time is much larger than the momentum life-time, an electron wanders over large portions of the Fermi-surface before spin-relaxation occurs, i.e. the hot-spots dominate the spin-relaxation. This effect is pronounced for metals where the Fermi surface strongly deviates from a sphere. We speculate that the deviation observed for BDD from the linear scaling is also caused by a similar effect but its verification requires additional theoretical work.

\section{Conclusions}
In summary, we identified the ESR signal of conduction electrons in boron doped superconducting diamond. The identification is based on the temperature dependence of the ESR signal intensity and its absolute magnitude. We find that the spin-relaxation mechanism in BDD is dominated by the Elliott-Yafet mechanism. However, we observe an anomalous relationship between the $g$-factor and the spin-relaxation time, which calls for further theoretical studies. The observed spin-relaxation rate is orders of smaller than the conventional theory predicts, which enhances the application potential of boron doped diamond for spintronics.

\section{Acknowledgements}
Useful discussions with Andr\'{a}s J\'{a}nossy are acknowledged. Work supported by the ERC Grant Nr.~ERC-259374-Sylo, and by the New Sz\'{e}chenyi Plan Nr.~T\'{A}MOP-4.2.2.B-10/1.2010-0009, and by the Hungarian State Grants (OTKA) Nr.~K81492. The Swiss NSF is acknowledged for support.

\onecolumngrid


\section*{SUPPLEMENTARY MATERIAL}
\setcounter{section}{0}
\setcounter{figure}{0}
\makeatletter 
\renewcommand{\thefigure}{S\@arabic\c@figure} 

\section{SEM and transport measurements}

\begin{figure}[!ht]

\begin{minipage}[b]{0.5\linewidth}
\centering
\includegraphics*[width=0.9\linewidth,height=0.7\linewidth]{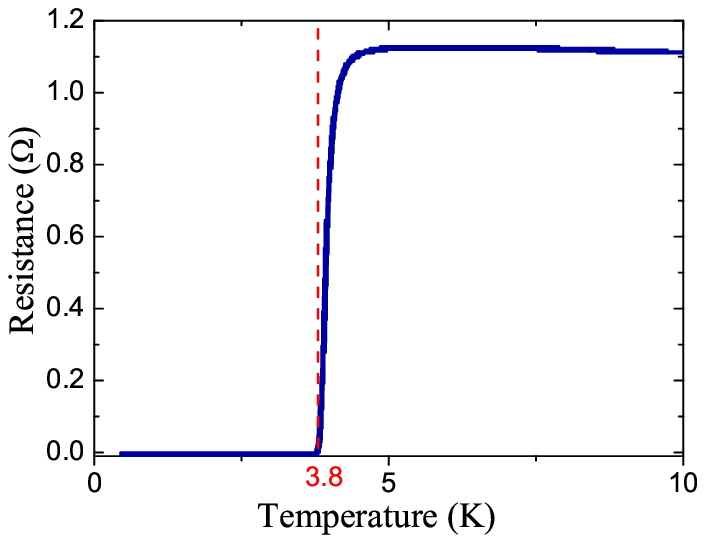}
\caption{Transport measurements indicate superconductivity below $T_{\text{c}}\approx 3.8~\text{K}$.}
\label{resistance}
\end{minipage}
\hspace{0.5cm}
\begin{minipage}[b]{0.46\linewidth}

\centering
\includegraphics*[width=\linewidth]{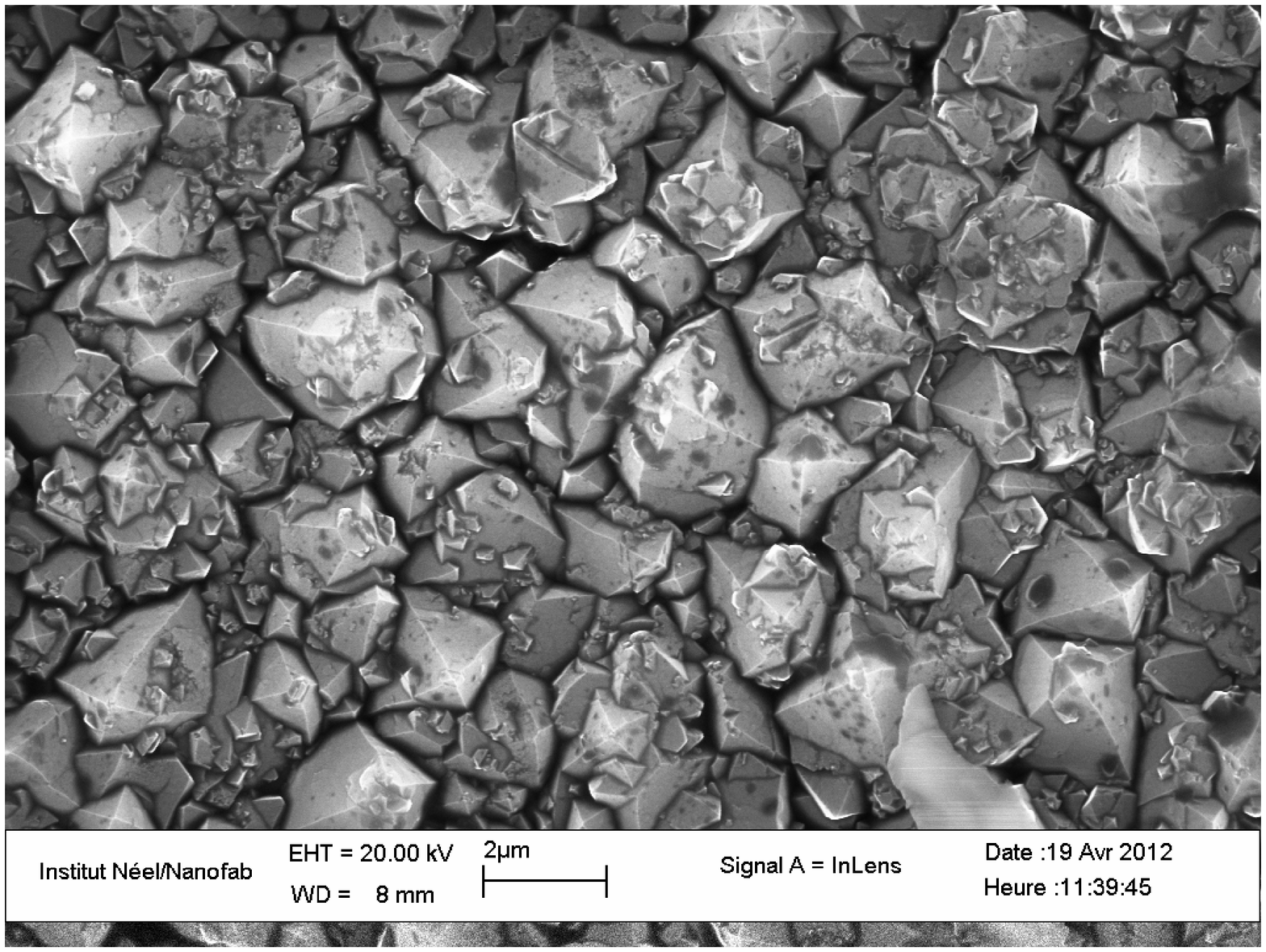}
\caption{Scanning Electron Microscope (SEM) image of the surface of the sample, consisting of grains of typically $3~\mu\text{m}$ size.}

\label{SEM}
\end{minipage}
\end{figure}

Fig.~\ref{resistance}.\ shows transport measurements in our sample. The spectrum exhibits the onset of superconductivity at 3.8~K. Superconducting properties of similar nanocrystals of boron doped diamond are analyzed in Ref.~\onlinecite{BDD_Mandal2010}.\\
Fig.~\ref{SEM}.\ depicts the SEM image of our material. The average diameter of the grains is $3~\mu\text{m}$, which is consistent with the nucleation density ($\gtrapprox10^{11}~\text{cm}^{-2}$).

\section{Raman measurements}
\begin{figure}[!tb]%
\includegraphics*[width=0.5\linewidth]{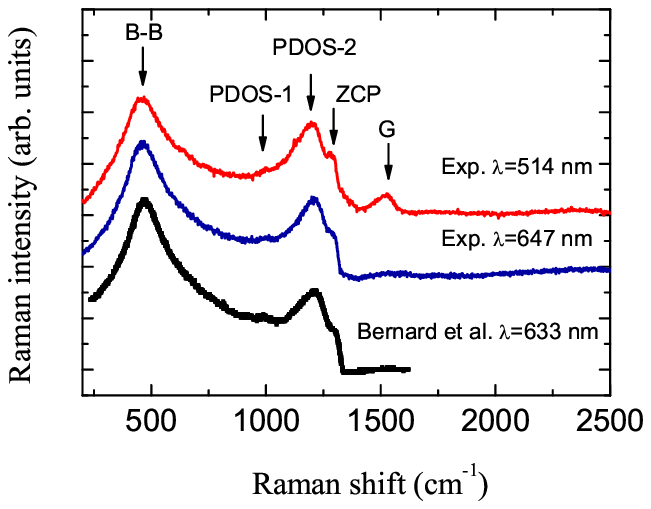}
\caption{Raman spectra of BDD at $\lambda=514~\text{nm}$ (upper curve) and at $\lambda=647~\text{nm}$ (middle curve). For comparison, we show Raman spectrum of an $n=6000~\text{ppm}$ polycrystalline BDD film from Ref.~\onlinecite{Bernard2004}. Labels denote the boron dimers (B-B), peaks due to the maxima of the phonon DOS (PDOS-1, PDOS-2), the mode related to the zone center phonon of diamond (ZCP), and the G band of \textit{sp$^2$} carbon (G).}

\label{Raman}
\end{figure}
In Fig.~\ref{Raman}., we show Raman spectra of our sample at two different wavelengths. As a comparison, we plotted the Raman spectrum from Ref.~\onlinecite{Bernard2004}. We identify similar bands in our spectra.\\
At $\sim 500~\text{cm}^{-1}$ (denoted by B-B in Fig.~\ref{Raman}.), the Raman band was assigned to boron dimers~\cite{BDD_Bourgeois, BDD_Bernard2004_DandRelMat, BDD_Sidorov2010} and to clustered boron atoms.~\cite{BDD_Sidorov2010} Note that the low energy of the phonon mode reflects weak force constant between boron atoms.
The peak was fitted with the sum of a Lorentzian and a Gaussian component.~\cite{BDD_Bernard2004_DandRelMat} The empirical relationship
\begin{equation}
n[\text{cm}^{-3}]=8.44\cdot 10^{30}\cdot \text{exp}\left(-0.048w[\text{cm}^{-1}]\right)
\label{Raman_conc}
\end{equation}
 was found between the wavenumber of the Lorentzian component ($w$) and the boron content ($n$) measured by secondary ion mass spectrometry (SIMS).~\cite{BDD_Bernard2004_DandRelMat} Using this equation, we obtain $n\approx 1.8\cdot 10^{21}~\text{cm}^{-3}$ for the boron content in our sample.\\

The Raman structure around $1200~\text{cm}^{-1}$ consists of two lines: a Lorentzian like around $1225~\text{cm}^{-1}$ and another one with an asymmetric lineshape around $1285~\text{cm}^{-1}$. \\
The zone center optical phonon of diamond, which appears at $1332~\text{cm}^{-1}$, is shifted to $1285~\text{cm}^{-1}$ (ZCP) in BDD and acquires a Fano lineshape due to the presence of free charge carriers.~\cite{BDD_Pruvost}\\
The Raman bands around $1000~\text{cm}^{-1}$ (PDOS-1) and $1225~\text{cm}^{-1}$ (PDOS-2) appear due to the presence of defects. These make the otherwise forbidden states, which belong to the maxima of the phonon density of states, allowed.~\cite{BDD_Sidorov2010, BDD_Vlasov}\\ The Raman band around $1520~\text{cm}^{-1}$ is assigned to the G band, and is downshifted upon boron doping.~\cite{BDD_Gajewski} The occurence of this Raman band points to the presence of \textit{sp$^2$} carbon bonds in the system.~\cite{FerrariPRB2000, FerrariPRB2001}\\

\section{Density of states of BDD}
\begin{figure}[tb]

\begin{minipage}[b]{0.48\linewidth}
\centering
\includegraphics*[width=0.9\linewidth,height=1.26\linewidth]{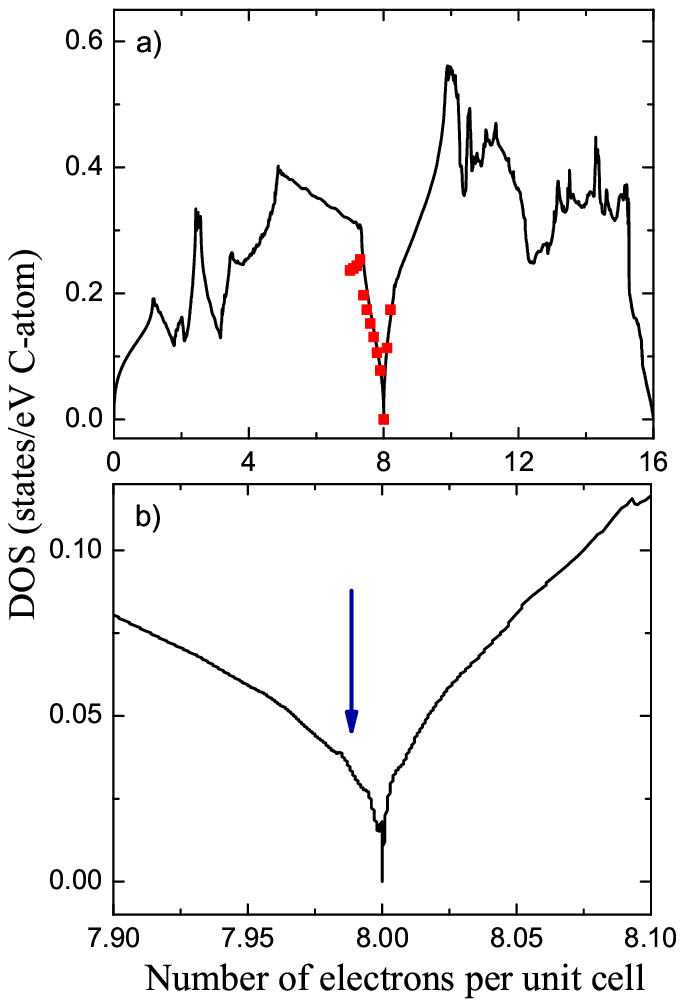}
\caption{Density of states calculated with DFT methods. $a)$ Density of states of neutral diamond as a function of number of electrons per unit cell (solid line) within LDA approximation. Density of states with optimized geometry upon adding extra charges to the system (\textcolor{red}{$\blacksquare$}) $b)$ Zoom on the calculated DOS for neutral diamond. The charge state for our BDD sample is labeled by arrow at 7.9872.}
\label{dosdft}
\end{minipage}
\hspace{0.5cm}
\begin{minipage}[b]{0.48\linewidth}

\centering
\includegraphics*[width=0.9\linewidth,height=1.3\linewidth]{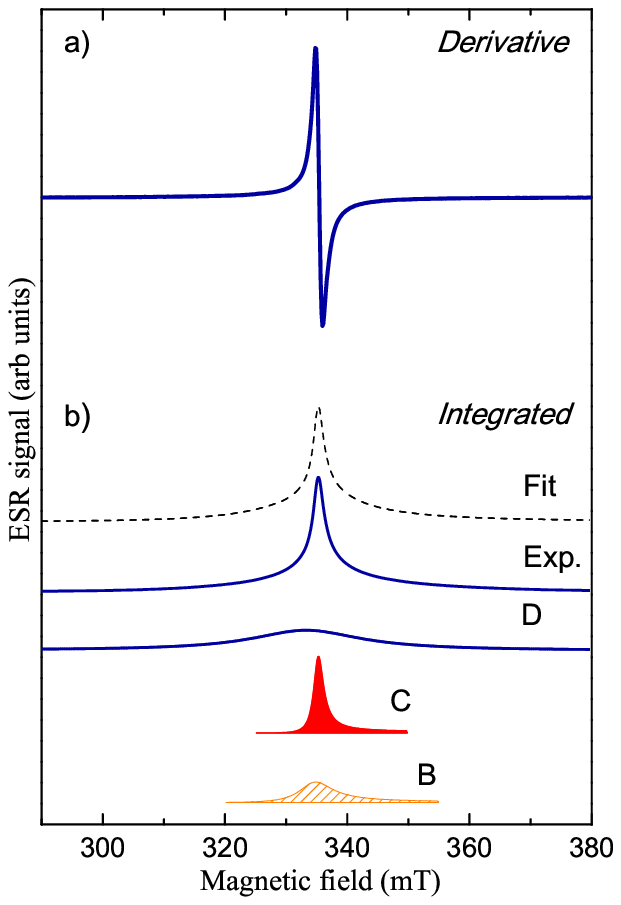}
\caption{\textit{a)} The measured derivative ESR spectrum at 35~K. \textit{b)} The integrated ESR spectrum (Exp.) and a fit (Fit), and the fitted components. \textit{B}, \textit{C} and \textit{D} lines probably originate from carbon dangling bonds accompanying hydrogen-vacancy complexes. Note that the \textit{C} signal is slightly asymmetric because of the finite penetration depth.}
\label{ESRspectrum35K}
\end{minipage}
\end{figure}

The density of states in the Fermi-gas model only depends on the free carrier concentration ($n$) through the Fermi wavenumber: $k_{\text{F}}=(3 \pi^2 n)^{1/3}$. The corresponding Fermi energy is $E_{{\text{F}}}=\hbar^2 k_{\text{F}}^2/2m^*$, where $m^*$ is the band effective mass. The present carrier density corresponds to a Fermi energy of $E_{{\textrm F}}=0.39~\text{eV}$. The DOS at the Fermi energy reads:

\begin{gather}
D(E_{\text{F}})=\frac{1}{2\pi^2}\left(\frac{2 m^*}{\hbar^2}\right)^{3/2}\sqrt{E_{\text{F}}}\frac{1}{\varrho},
\label{BDD_DOS}
\end{gather}

\noindent in units of $\text{states}/\text{eV}\cdot \text{C-atom} $, where $\varrho=1.76\,\cdot 10^{23}~\text{ C-atom}/\text{cm}^3$ is the atomic density of diamond.
For the effective mass of charge carriers, the free electron mass is substituted.~\cite{BDD_Nature2004} This gives $D(E_{\textrm F})=2.4\cdot 10^{-2}~\text{states}/\text{eV}\cdot\text{C-atom}$.\\

In order to calculate the density of states from angle-resolved photoemission spectroscopy (ARPES) measurements,~\cite{BDD_Nature2005} the Fermi surface is regarded. Within the Fermi-gas model, the density of states can be rewritten as
\begin{equation}
D(E_{\text{F}})=\frac{1}{\pi^2}\frac{k_{\text{F}}^2}{v_{\text{F}}\hbar}\frac{1}{\varrho},
\label{BDD_DOS2}
\end{equation}
where $v_{\text{F}}$ is the Fermi velocity. The given values of $v_{\text{F}}$ and $k_{\text{F}}$ from Ref.~\onlinecite{BDD_Nature2005} for 'BDD2' ($T_{\text{c}}=2.5~\text{K}$) and 'BDD3' ($T_{\text{c}}=7~\text{K}$) samples give $D(E_{\text{F}})=3.9\cdot 10^{-3}~\text{states}/(\text{eV}\cdot\text{C-atom})$ and $D(E_{\text{F}})=7.8\cdot 10^{-3}~\text{states}/(\text{eV}\cdot\text{C-atom})$.\\

We performed density functional theory calculations (DFT) with the Vienna ab initio Simulation Package (VASP)~\cite{KresseG_1996_2} within the local density approximation (LDA) to calculate the density of states of neutral diamond (see Fig.~\ref{dosdft}). The projector augmented-wave method was used with a plane-wave cutoff energy of 750~eV and a k-point set of 30$\times$30$\times$30 $\Gamma$-centered Monkhorst-Pack grid.\\
A series of geometry optimization and density of states calculation with extra charges added (removed) from the system were also performed. The results are shown in Fig.~\ref{dosdft}a. The agreement between the DOS calculated for neutral and charged diamond firmly supports the validity of the rigid band approximation at low doping levels. The boron content of 6400~ppm yields $D(E_\text{F})=3.6\cdot 10^{-2}~\text{states}/(\text{eV}\cdot\text{C-atom})$ for the DOS.\\

\section{ESR measurements at low temperature}

In Fig.~\ref{ESRspectrum35K}a., we show a typical ESR spectrum of BDD at low temperature (35~K). While on the derivative spectrum less visible, the integrated spectrum in Fig.~\ref{ESRspectrum35K}b.\ shows the presence of the paramagnetic \textit{D} signal together with the \textit{B} and \textit{C} signals. We showed in the main text that \textit{B} and \textit{C} are also present at high temperature. However, the signal \textit{D} cannot be resolved above 75~K.\\

In the ESR spectrum of BDD, we identified three different paramagnetic impurities. The intensities of \textit{B} and \textit{C} are similar. Intensity calibration gives their spin concentration to be 0.6~ppm ($n_{\text{imp}(C)}\approx n_{\text{imp}(B)}\approx 10^{17}~\text{cm}^{-3}$). The average distance of paramagnetic spin-1/2 impurities is about 14~nm. The intensity of \textit{D} is three times larger, which corresponds to 1.8~ppm spin concentration ($n_{\text{imp}(D)}\approx 3\cdot 10^{17}~\text{cm}^{-3}$), and about 10~nm average distance. The spin concentration of the defects is in good agreement with bulk carbon dangling bond defects accompanying hydrogen-vacancy complexes found both in weakly boron-doped and as-grown CVD diamond.~\cite{BDD_Zhou, BDD_Mizuochi1, BDD_Mizuochi2, BDD_Mizuochi3}\\
This accordance points to the fact that the quantity of these bulk defects mainly depends on the CH$_4$/H$_2$ ratio used for the growth of BDD.~\cite{BDD_Mizuochi3} The CH$_4$/H$_2$ ratio controls the grain size~\cite{BDD_Janssens} and modifies the surface-to-bulk ratio. As the hydrogen concentration is higher on the grain boundaries,~\cite{BDD_Liao} the defect concentration is expected to have a maximum there. The high concentration of defects within the penetration depth explains the asymmetry of the \textit{C} signal at low temperature.\\

\section{ESR intensity calibration}
\label{calibration}

ESR spectroscopy measures the net amount of magnetic moments, which is proportional to the sample amount. In order to gain information on the corresponding intensive variable, the spin-susceptibility ($\chi_s$), the ESR intensity of an unknown material must be calibrated against a Curie (i.\ e.\ non-interacting) spin system with known amount of spins.~\cite{SzirmaiPSSB2011}\\
The molar Curie spin-susceptibility reads
\begin{equation}
\chi_s(\text{Curie})=\frac{g^2S(S+1)\mu_{\text{B}}^2}{3k_{\text{B}}T}\cdot N_{\text{A}},
\label{Curie}
\end{equation}
where $S$ denotes the spin state of the non-interacting spins, $\mu_{\text{B}}$ is the Bohr moment, $k_{\text{B}}$ is the Boltzmann constant, and $N_\text{A}$ is the Avogadro number. As an example, if we consider an $S=1/2$ system at 300~K and substitute the corresponding physical constants in Gaussian units into Eq.~\eqref{Curie}, we obtain the well-known $\chi_s(\text{Curie})=1.25\cdot 10^{-3}~\text{emu/mol}$.\\
The Pauli spin-susceptibility for itinerant electrons in a metal is
\begin{equation}
\chi_s(\text{Pauli})=\frac{g^2\mu_{\text{B}}^2}{4}D(E_{\text{F}})\cdot N_{\text{A}}.
\label{Pauli}
\end{equation}
The spin-susceptibilities are related to the ESR intensity as
\begin{equation}
I_{\text{ESR}}\propto\sum m=B_{\text{res}}\chi_s\cdot n,
\label{intensity}
\end{equation}
where $n$ is the amount of the sample and $B_{\text{res}}$ is the magnetic field of the resonance. Therefore, the comparison of the ESR intensity of an unknown material with itinerant electrons to a known amount of Curie spins yields (for $S=1/2$ and $g_{\text{Pauli}},g_{\text{Curie}}\approx 2$)
\begin{equation}
\frac{I_{\text{ESR}}(\text{Pauli})}{I_{\text{ESR}}(\text{Curie})}=k_{\text{B}}T D(E_{\text{F}})\frac{n(\text{Pauli})}{n(\text{Curie})}.
\label{comparison}
\end{equation}

Eq.~\eqref{comparison} can be used to determine the DOS at the Fermi energy, $D(E_{\text{F}})$.\\

\section{The ESR lineshape}
In general, the ESR signal can be expressed as a sum of dispersion ($f^{\text{disp}}$) and absorption ($f^{\text{abs}}$) Lorentzian lines:
\begin{equation}
P \propto \cos{\varphi}\cdot f^{\text{abs}}(B-B_0,w)+\sin{\varphi}\cdot f^{\text{disp}}(B-B_0,w),
\label{lineshape}
\end{equation}
where $B$ is the magnetic field, $B_0$ is the resonance magnetic field, and $w$ is the linewidth.\\
The \textit{A} line is best fitted with $\varphi\approx45^{\circ}$ in the temperature range of 75~K to 300~K. This phase describes the ESR lineshape of metals in the thick plate case.~\cite{FeherKip}\\
The conditions for the validity of this case are $T_T\gg T_D$ and $T_D\gg T_1$, where $T_T$ is the time it takes for an electron to diffusively traverse the sample and $T_D$ is the time it takes to diffuse through the skin depth.\\
$T_D$ is defined by $\delta=v_{\text{F}}\sqrt{\tau T_D/d}$, where $\delta$ is the penetration depth, $v_{\text{F}}$ is the Fermi velocity, $\tau$ is the momentum relaxation time, and $d=3$ is the dimensionality. With the spin-diffusion length of $\delta_{\text{spin}}=v_{\text{F}}\sqrt{\tau T_1/d}$, the $T_D\gg T_1$ condition is equivalent to $\delta\gg \delta_{\text{spin}}$.\\
The microwave penetration depth reads $\delta=\sqrt{2/\mu_0 \omega \sigma}$, where $\mu_0$ is the permeability of the vacuum, $\omega$ is the angular frequency of the exciting microwave, and $\sigma$ is the conductivity. With room-temperature resistivity data in nanocrystalline diamond from Ref.~\onlinecite{BDD_Mares}, the penetration depth is estimated\cite{Sigma} to be $\delta\approx 31~\mu\text{m}$. At low temperature, $\delta\approx 33~\mu\text{m}$ is calculated. From the ESR linewidth, $T_1\approx 0.2~\text{ns}$. With the Fermi velocity ($v_{\text{F}}(\text{BDD2})\approx 1.1\cdot 10^6~\text{m/s}$) and momentum relaxation time ($\tau(\text{BDD2})=5.1~\text{fs}$) from Ref.~\onlinecite{BDD_Nature2005}, the spin-diffusion length is $\delta_{\text{spin}}(\text{BDD2})\approx 0.6~\mu\text{m}$. (Note that the momentum relaxation time from the Drude model yields a similar value.) These estimated values show that the condition $\delta\gg \delta_{\text{spin}}$ holds.\\

\section{Measurement of the \lowercase{\textit{g}}-factor}
\label{gfactor}
The Zeeman level energy splitting and the microwave excitation energy are related by

\begin{equation}
\hbar\omega=g\mu_{\text{B}}B_{\text{res}},
\label{zeeman}
\end{equation}
where $\omega$ is the transition angular frequency, and $B_{\text{res}}$ is the resonance magnetic field. Typical values in X-band ESR are ($\sim~9.4~\text{GHz}$): $\omega\approx5.9\cdot 10^{10}~\text{rad/s}$ and $B_{\text{res}}\approx 330~\text{mT}$, for $g\approx2$. This is close to the free electron $g$-factor ($g\approx2.0023$). In principle, an independent measurement of the microwave frequency and the resonance magnetic field yields the $g$-factor for an unknown material. This is however impractical due to the uncertainty of the magnetic field measurement at the sample. An alternative, which yields accurate $g$-factor values is to measure the unknown sample together with a reference sample with a well-known $g$-factor value. E.\ g.\ $\text{Mn}^{\text{2+}}\text{:MgO}$ ($g=2.0014$ Ref.~\onlinecite{AbragamBleaneyBook}) is a usual choice as it does not have lines at $g=2$ and the hyperfine interaction can be accounted for easily until the third order. Using a mixture of the BDD sample and $\text{Mn}^{\text{2+}}\text{:MgO}$ gives $\omega(\text{BDD})=\omega(\text{Mn}^{\text{2+}}\text{:MgO})$, and simplifies Eq.~\eqref{zeeman} to

\begin{equation}
\frac{g(\text{BDD})}{g(\text{Mn}^{\text{2+}}\text{:MgO})}=\frac{B_{\text{res}}(\text{Mn}^{\text{2+}}\text{:MgO})}{B_{\text{res}}(\text{BDD})}.
\label{zeeman2}
\end{equation}

The resonance field for BDD and $\text{Mn}^{\text{2+}}\text{:MgO}$ is obtained by simultaneously fitting derivative Lorentzian lines to the measured spectrum (by taking into account the second order hyperfine interaction of $\text{Mn}^{\text{2+}}$.

\section{Calibration of the \lowercase{\textit{g}}-factor and the ESR intensity in BDD}

\begin{figure}[!tb]%
\includegraphics*[width=0.48\linewidth]{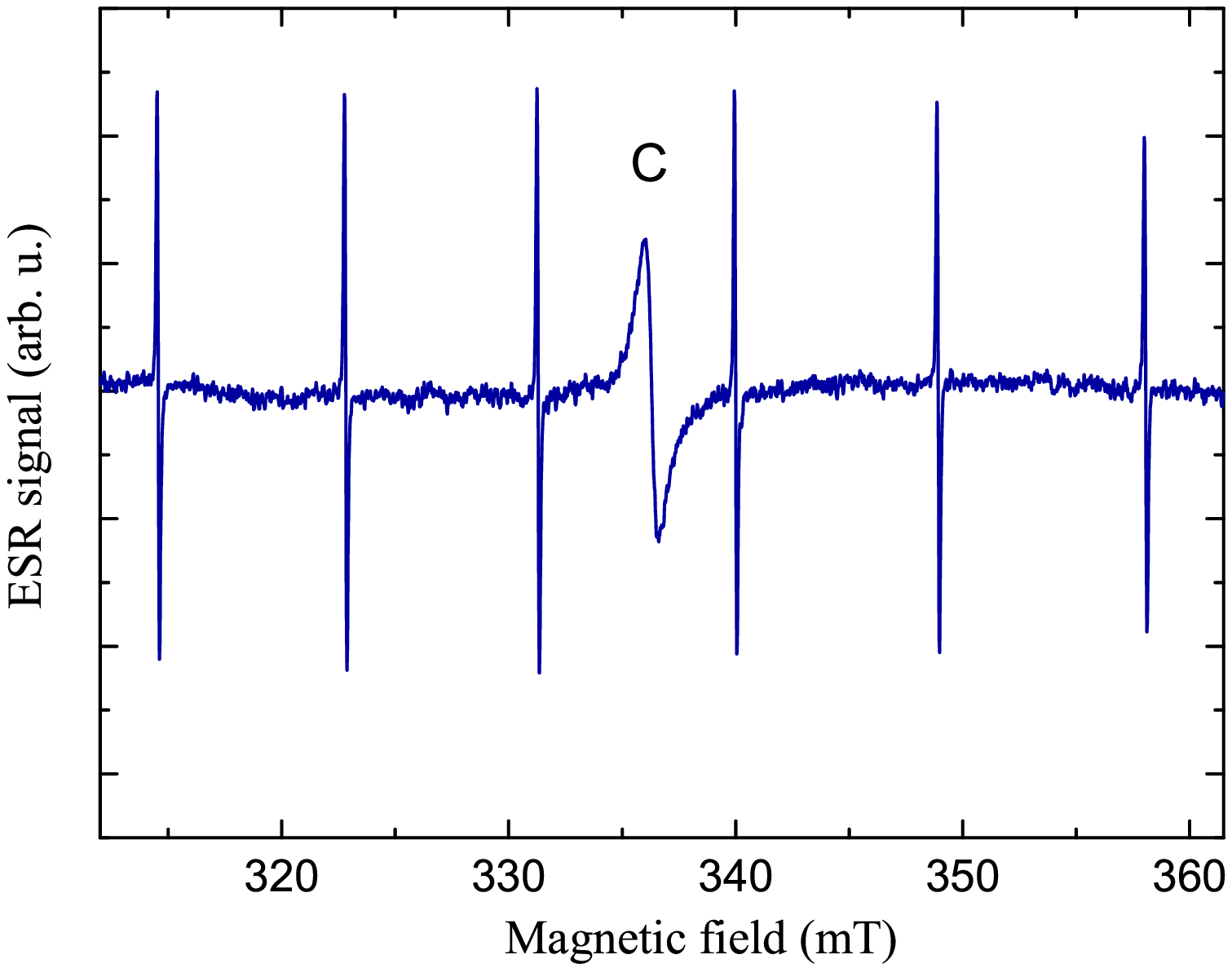}
\caption{ESR spectrum of a mixture of BDD and $\text{Mn}^{\text{2+}}\text{:MgO}$ powder at room temperature. The closely equidistant sextuplet comes from the $\left|-1/2\right\rangle \rightarrow \left|1/2\right\rangle$ Zeeman transition of $\text{Mn}^{\text{2+}}$. The broader central line is the \textit{C} signal in BDD. (The modulation amplitude is $50~\mu\text{T}$, the microwave power is 1~mW.)}

\label{MnBDD}
\end{figure}

In Fig.~\ref{MnBDD}., we show the ESR spectrum of mixed $\text{Mn}^{\text{2+}}\text{:MgO}$ and BDD powder. The spectrum is a superposition of the sextuplet of $\text{Mn}^{\text{2+}}\text{:MgO}$ ($B_{\text{res}}(\text{Mn}^{\text{2+}}\text{:MgO})=336.6~\text{mT}$) and the \textit{C} signal of BDD ($B_{\text{res}}=336.3~\text{mT}$).\\
Care was taken to employ low modulation ($50~\mu\text{T}$) and low microwave power (1~mW). Note that this modulation amplitude is too low for the measurement of broader impurity-related components (\textit{B} and \textit{D}) and especially for the CESR signal (\textit{A}). Other spectra (shown in the Supplementary Material and in the main text) are therefore measured with higher magnetic field modulation amplitude (0.5~mT as the highest) in order to increase the signal to noise ratio. Measurements carried out with the highest modulation amplitude overmodulate the signal \textit{C}.\\
As the ESR signal of the standard ($\text{Mn}^{\text{2+}}\text{:MgO}$) and the \textit{A} signal cannot be resolved simultaneously, the DOS in BDD is thus determined indirectly. Once the spin-susceptibility of the \textit{C} signal is calculated in the mixture of $\text{Mn}^{\text{2+}}\text{:MgO}$ and BDD, the comparison of the ESR signal intensity of \textit{C} (fitted with the overmodulated lineshape) and that of \textit{A} gives the DOS. The spin-susceptibility of \textit{B} and \textit{D} is determined similarly.\\
The $g$-factor calibration, described in the previous section is performed using the same indirect method, i.e.\ first calibrating the $g$-factor for the $C$ signal when using low magnetic field modulation amplitude and then using it as a reference to calibrate the corresponding values for the other signals ($A$, $B$, and $D$).\\

\twocolumngrid


\begin{thebibliography}{53}
\expandafter\ifx\csname natexlab\endcsname\relax\def\natexlab#1{#1}\fi
\expandafter\ifx\csname bibnamefont\endcsname\relax
  \def\bibnamefont#1{#1}\fi
\expandafter\ifx\csname bibfnamefont\endcsname\relax
  \def\bibfnamefont#1{#1}\fi
\expandafter\ifx\csname citenamefont\endcsname\relax
  \def\citenamefont#1{#1}\fi
\expandafter\ifx\csname url\endcsname\relax
  \def\url#1{\texttt{#1}}\fi
\expandafter\ifx\csname urlprefix\endcsname\relax\def\urlprefix{URL }\fi
\providecommand{\bibinfo}[2]{#2}
\providecommand{\eprint}[2][]{\url{#2}}

\bibitem[{\citenamefont{\v{Z}uti\'c et~al.}(2004)\citenamefont{\v{Z}uti\'c,
  Fabian, and Sarma}}]{FabianRMP}
\bibinfo{author}{\bibfnamefont{I.}~\bibnamefont{\v{Z}uti\'c}},
  \bibinfo{author}{\bibfnamefont{J.}~\bibnamefont{Fabian}}, \bibnamefont{and}
  \bibinfo{author}{\bibfnamefont{S.~D.} \bibnamefont{Sarma}},
  \bibinfo{journal}{Rev. Mod. Phys.} \textbf{\bibinfo{volume}{76}},
  \bibinfo{pages}{323} (\bibinfo{year}{2004}).

\bibitem[{\citenamefont{Wu et~al.}(2010)\citenamefont{Wu, Jiang, and
  Weng}}]{WuReview}
\bibinfo{author}{\bibfnamefont{M.}~\bibnamefont{Wu}},
  \bibinfo{author}{\bibfnamefont{J.}~\bibnamefont{Jiang}}, \bibnamefont{and}
  \bibinfo{author}{\bibfnamefont{M.}~\bibnamefont{Weng}},
  \bibinfo{journal}{Phys. Rep.} \textbf{\bibinfo{volume}{493}},
  \bibinfo{pages}{61 } (\bibinfo{year}{2010}).

\bibitem[{\citenamefont{Tombros et~al.}(2007)\citenamefont{Tombros, J\'{o}zsa,
  Popinciuc, Jonkman, and van Wees}}]{TombrosNAT2007}
\bibinfo{author}{\bibfnamefont{N.}~\bibnamefont{Tombros}},
  \bibinfo{author}{\bibfnamefont{C.}~\bibnamefont{J\'{o}zsa}},
  \bibinfo{author}{\bibfnamefont{M.}~\bibnamefont{Popinciuc}},
  \bibinfo{author}{\bibfnamefont{H.~T.} \bibnamefont{Jonkman}},
  \bibnamefont{and} \bibinfo{author}{\bibfnamefont{B.~J.} \bibnamefont{van
  Wees}}, \bibinfo{journal}{Nature} \textbf{\bibinfo{volume}{448}},
  \bibinfo{pages}{571} (\bibinfo{year}{2007}).

\bibitem[{\citenamefont{Yang et~al.}(2011)\citenamefont{Yang, Balakrishnan,
  Volmer, Avsar, Jaiswal, Samm, Ali, Pachoud, Zeng, Popinciuc
  et~al.}}]{GuntherodtBilayer}
\bibinfo{author}{\bibfnamefont{T.-Y.} \bibnamefont{Yang}},
  \bibinfo{author}{\bibfnamefont{J.}~\bibnamefont{Balakrishnan}},
  \bibinfo{author}{\bibfnamefont{F.}~\bibnamefont{Volmer}},
  \bibinfo{author}{\bibfnamefont{A.}~\bibnamefont{Avsar}},
  \bibinfo{author}{\bibfnamefont{M.}~\bibnamefont{Jaiswal}},
  \bibinfo{author}{\bibfnamefont{J.}~\bibnamefont{Samm}},
  \bibinfo{author}{\bibfnamefont{S.~R.} \bibnamefont{Ali}},
  \bibinfo{author}{\bibfnamefont{A.}~\bibnamefont{Pachoud}},
  \bibinfo{author}{\bibfnamefont{M.}~\bibnamefont{Zeng}},
  \bibinfo{author}{\bibfnamefont{M.}~\bibnamefont{Popinciuc}},
  \bibnamefont{et~al.}, \bibinfo{journal}{Phys. Rev. Lett.}
  \textbf{\bibinfo{volume}{107}}, \bibinfo{pages}{047206}
  (\bibinfo{year}{2011}).

\bibitem[{\citenamefont{Han and Kawakami}(2011)}]{KawakamiBilayer}
\bibinfo{author}{\bibfnamefont{W.}~\bibnamefont{Han}} \bibnamefont{and}
  \bibinfo{author}{\bibfnamefont{R.~K.} \bibnamefont{Kawakami}},
  \bibinfo{journal}{Phys. Rev. Lett.} \textbf{\bibinfo{volume}{107}},
  \bibinfo{pages}{047207} (\bibinfo{year}{2011}).

\bibitem[{\citenamefont{Griswold et~al.}(1952)\citenamefont{Griswold, Kip, and
  Kittel}}]{KipKittelPR1952}
\bibinfo{author}{\bibfnamefont{T.~W.} \bibnamefont{Griswold}},
  \bibinfo{author}{\bibfnamefont{A.~F.} \bibnamefont{Kip}}, \bibnamefont{and}
  \bibinfo{author}{\bibfnamefont{C.}~\bibnamefont{Kittel}},
  \bibinfo{journal}{Phys. Rev.} \textbf{\bibinfo{volume}{88}},
  \bibinfo{pages}{951} (\bibinfo{year}{1952}).

\bibitem[{\citenamefont{Mott}(1968)}]{Mott1968}
\bibinfo{author}{\bibfnamefont{N.}~\bibnamefont{Mott}}, \bibinfo{journal}{J.
  Non-Cryst. Solids} \textbf{\bibinfo{volume}{1}}, \bibinfo{pages}{1 }
  (\bibinfo{year}{1968}).

\bibitem[{\citenamefont{Klein et~al.}(2007)\citenamefont{Klein, Achatz,
  Kacmarcik, Marcenat, Gustafsson, Marcus, Bustarret, Pernot, Omnes, Sernelius
  et~al.}}]{BDD_KleinPRB}
\bibinfo{author}{\bibfnamefont{T.}~\bibnamefont{Klein}},
  \bibinfo{author}{\bibfnamefont{P.}~\bibnamefont{Achatz}},
  \bibinfo{author}{\bibfnamefont{J.}~\bibnamefont{Kacmarcik}},
  \bibinfo{author}{\bibfnamefont{C.}~\bibnamefont{Marcenat}},
  \bibinfo{author}{\bibfnamefont{F.}~\bibnamefont{Gustafsson}},
  \bibinfo{author}{\bibfnamefont{J.}~\bibnamefont{Marcus}},
  \bibinfo{author}{\bibfnamefont{E.}~\bibnamefont{Bustarret}},
  \bibinfo{author}{\bibfnamefont{J.}~\bibnamefont{Pernot}},
  \bibinfo{author}{\bibfnamefont{F.}~\bibnamefont{Omnes}},
  \bibinfo{author}{\bibfnamefont{B.~E.} \bibnamefont{Sernelius}},
  \bibnamefont{et~al.}, \bibinfo{journal}{Phys. Rev. B}
  \textbf{\bibinfo{volume}{75}}, \bibinfo{pages}{165313}
  (\bibinfo{year}{2007}).

\bibitem[{\citenamefont{Bustarret et~al.}(2008)\citenamefont{Bustarret, Achatz,
  Sac\'{e}p\'{e}, Chapelier, Marcenat, Ort\'{e}ga, and
  Klein}}]{BDD_Bustarret2008}
\bibinfo{author}{\bibfnamefont{E.}~\bibnamefont{Bustarret}},
  \bibinfo{author}{\bibfnamefont{P.}~\bibnamefont{Achatz}},
  \bibinfo{author}{\bibfnamefont{B.}~\bibnamefont{Sac\'{e}p\'{e}}},
  \bibinfo{author}{\bibfnamefont{C.}~\bibnamefont{Chapelier}},
  \bibinfo{author}{\bibfnamefont{C.}~\bibnamefont{Marcenat}},
  \bibinfo{author}{\bibfnamefont{L.}~\bibnamefont{Ort\'{e}ga}},
  \bibnamefont{and} \bibinfo{author}{\bibfnamefont{T.}~\bibnamefont{Klein}},
  \bibinfo{journal}{Philos. T. Roy. Soc. A} \textbf{\bibinfo{volume}{366}},
  \bibinfo{pages}{267} (\bibinfo{year}{2008}).

\bibitem[{Car()}]{Carrier}
\bibinfo{note}{We refer to the nominal concentration of the boron atoms
  throughout the text which is higher than the actual carrier concentration due
  to the tendency of boron dimer formation which does not contribute to free
  carriers.}

\bibitem[{\citenamefont{Ekimov et~al.}(2004)\citenamefont{Ekimov, Sidorov,
  Bauer, Mel'nik, Curro, Thompson, and Stishov}}]{BDD_Nature2004}
\bibinfo{author}{\bibfnamefont{E.~A.} \bibnamefont{Ekimov}},
  \bibinfo{author}{\bibfnamefont{V.~A.} \bibnamefont{Sidorov}},
  \bibinfo{author}{\bibfnamefont{E.~D.} \bibnamefont{Bauer}},
  \bibinfo{author}{\bibfnamefont{N.~N.} \bibnamefont{Mel'nik}},
  \bibinfo{author}{\bibfnamefont{N.~J.} \bibnamefont{Curro}},
  \bibinfo{author}{\bibfnamefont{J.~D.} \bibnamefont{Thompson}},
  \bibnamefont{and} \bibinfo{author}{\bibfnamefont{S.~M.}
  \bibnamefont{Stishov}}, \bibinfo{journal}{Nature}
  \textbf{\bibinfo{volume}{428}}, \bibinfo{pages}{542} (\bibinfo{year}{2004}).

\bibitem[{\citenamefont{Yokoya et~al.}(2005)\citenamefont{Yokoya, Nakamura,
  Matsushita, Muro, Takano, Nagao, Takenouchi, Kawarada, and
  Oguchi}}]{BDD_Nature2005}
\bibinfo{author}{\bibfnamefont{T.}~\bibnamefont{Yokoya}},
  \bibinfo{author}{\bibfnamefont{T.}~\bibnamefont{Nakamura}},
  \bibinfo{author}{\bibfnamefont{T.}~\bibnamefont{Matsushita}},
  \bibinfo{author}{\bibfnamefont{T.}~\bibnamefont{Muro}},
  \bibinfo{author}{\bibfnamefont{Y.}~\bibnamefont{Takano}},
  \bibinfo{author}{\bibfnamefont{M.}~\bibnamefont{Nagao}},
  \bibinfo{author}{\bibfnamefont{T.}~\bibnamefont{Takenouchi}},
  \bibinfo{author}{\bibfnamefont{H.}~\bibnamefont{Kawarada}}, \bibnamefont{and}
  \bibinfo{author}{\bibfnamefont{T.}~\bibnamefont{Oguchi}},
  \bibinfo{journal}{Nature} \textbf{\bibinfo{volume}{438}},
  \bibinfo{pages}{647} (\bibinfo{year}{2005}).

\bibitem[{\citenamefont{Mandal et~al.}(2011)\citenamefont{Mandal, Bautze,
  Williams, Naud, Bustarret, Omn\`{e}s, Rodi\`{e}re, Meunier, B\"{a}uerle, and
  Saminadayar}}]{BDD_Mandal}
\bibinfo{author}{\bibfnamefont{S.}~\bibnamefont{Mandal}},
  \bibinfo{author}{\bibfnamefont{T.}~\bibnamefont{Bautze}},
  \bibinfo{author}{\bibfnamefont{O.~A.} \bibnamefont{Williams}},
  \bibinfo{author}{\bibfnamefont{C.}~\bibnamefont{Naud}},
  \bibinfo{author}{\bibfnamefont{E.}~\bibnamefont{Bustarret}},
  \bibinfo{author}{\bibfnamefont{F.}~\bibnamefont{Omn\`{e}s}},
  \bibinfo{author}{\bibfnamefont{P.}~\bibnamefont{Rodi\`{e}re}},
  \bibinfo{author}{\bibfnamefont{T.}~\bibnamefont{Meunier}},
  \bibinfo{author}{\bibfnamefont{C.}~\bibnamefont{B\"{a}uerle}},
  \bibnamefont{and}
  \bibinfo{author}{\bibfnamefont{L.}~\bibnamefont{Saminadayar}},
  \bibinfo{journal}{ACS Nano} \textbf{\bibinfo{volume}{5}},
  \bibinfo{pages}{7144} (\bibinfo{year}{2011}).

\bibitem[{\citenamefont{Elliott}(1954)}]{Elliott}
\bibinfo{author}{\bibfnamefont{R.~J.} \bibnamefont{Elliott}},
  \bibinfo{journal}{Phys. Rev.} \textbf{\bibinfo{volume}{96}},
  \bibinfo{pages}{266} (\bibinfo{year}{1954}).

\bibitem[{\citenamefont{Yafet}(1983)}]{YafetPL1983}
\bibinfo{author}{\bibfnamefont{Y.}~\bibnamefont{Yafet}},
  \bibinfo{journal}{Phys. Lett. A} \textbf{\bibinfo{volume}{98}},
  \bibinfo{pages}{287} (\bibinfo{year}{1983}).

\bibitem[{\citenamefont{Beuneu and Monod}(1978)}]{BeuneuMonodPRB1978}
\bibinfo{author}{\bibfnamefont{F.}~\bibnamefont{Beuneu}} \bibnamefont{and}
  \bibinfo{author}{\bibfnamefont{P.}~\bibnamefont{Monod}},
  \bibinfo{journal}{Phys. Rev. B} \textbf{\bibinfo{volume}{18}},
  \bibinfo{pages}{2422} (\bibinfo{year}{1978}).

\bibitem[{\citenamefont{Hees et~al.}(2011)\citenamefont{Hees, Kriele, and
  Williams}}]{BDD_Hees}
\bibinfo{author}{\bibfnamefont{J.}~\bibnamefont{Hees}},
  \bibinfo{author}{\bibfnamefont{A.}~\bibnamefont{Kriele}}, \bibnamefont{and}
  \bibinfo{author}{\bibfnamefont{O.~A.} \bibnamefont{Williams}},
  \bibinfo{journal}{Chem. Phys. Lett.} \textbf{\bibinfo{volume}{509}},
  \bibinfo{pages}{12 } (\bibinfo{year}{2011}).

\bibitem[{\citenamefont{Gajewski et~al.}(2009)\citenamefont{Gajewski, Achatz,
  Williams, Haenen, Bustarret, Stutzmann, and Garrido}}]{BDD_Gajewski}
\bibinfo{author}{\bibfnamefont{W.}~\bibnamefont{Gajewski}},
  \bibinfo{author}{\bibfnamefont{P.}~\bibnamefont{Achatz}},
  \bibinfo{author}{\bibfnamefont{O.~A.} \bibnamefont{Williams}},
  \bibinfo{author}{\bibfnamefont{K.}~\bibnamefont{Haenen}},
  \bibinfo{author}{\bibfnamefont{E.}~\bibnamefont{Bustarret}},
  \bibinfo{author}{\bibfnamefont{M.}~\bibnamefont{Stutzmann}},
  \bibnamefont{and} \bibinfo{author}{\bibfnamefont{J.~A.}
  \bibnamefont{Garrido}}, \bibinfo{journal}{Phys. Rev. B}
  \textbf{\bibinfo{volume}{79}}, \bibinfo{pages}{045206}
  (\bibinfo{year}{2009}).

\bibitem[{Sup()}]{SupMat}
\bibinfo{note}{See supplementary material}.

\bibitem[{\citenamefont{Mukuda et~al.}(2006)\citenamefont{Mukuda, Tsuchida,
  Harada, Kitaoka, Takenouchi, Takano, Nagao, Sakaguchi, and
  Kawarada}}]{BDD_NMR_conc}
\bibinfo{author}{\bibfnamefont{H.}~\bibnamefont{Mukuda}},
  \bibinfo{author}{\bibfnamefont{T.}~\bibnamefont{Tsuchida}},
  \bibinfo{author}{\bibfnamefont{A.}~\bibnamefont{Harada}},
  \bibinfo{author}{\bibfnamefont{Y.}~\bibnamefont{Kitaoka}},
  \bibinfo{author}{\bibfnamefont{T.}~\bibnamefont{Takenouchi}},
  \bibinfo{author}{\bibfnamefont{Y.}~\bibnamefont{Takano}},
  \bibinfo{author}{\bibfnamefont{M.}~\bibnamefont{Nagao}},
  \bibinfo{author}{\bibfnamefont{I.}~\bibnamefont{Sakaguchi}},
  \bibnamefont{and} \bibinfo{author}{\bibfnamefont{H.}~\bibnamefont{Kawarada}},
  \bibinfo{journal}{Sci. Technol. Adv. Mat.} \textbf{\bibinfo{volume}{7}},
  \bibinfo{pages}{S37} (\bibinfo{year}{2006}).

\bibitem[{\citenamefont{Takano et~al.}(2007)\citenamefont{Takano, Takenouchi,
  Ishii, Ueda, Okutsu, Sakaguchi, Umezawa, Kawarada, and
  Tachiki}}]{BDD_CVD_Diamond2007}
\bibinfo{author}{\bibfnamefont{Y.}~\bibnamefont{Takano}},
  \bibinfo{author}{\bibfnamefont{T.}~\bibnamefont{Takenouchi}},
  \bibinfo{author}{\bibfnamefont{S.}~\bibnamefont{Ishii}},
  \bibinfo{author}{\bibfnamefont{S.}~\bibnamefont{Ueda}},
  \bibinfo{author}{\bibfnamefont{T.}~\bibnamefont{Okutsu}},
  \bibinfo{author}{\bibfnamefont{I.}~\bibnamefont{Sakaguchi}},
  \bibinfo{author}{\bibfnamefont{H.}~\bibnamefont{Umezawa}},
  \bibinfo{author}{\bibfnamefont{H.}~\bibnamefont{Kawarada}}, \bibnamefont{and}
  \bibinfo{author}{\bibfnamefont{M.}~\bibnamefont{Tachiki}},
  \bibinfo{journal}{Diam. Relat. Mater.} \textbf{\bibinfo{volume}{16}},
  \bibinfo{pages}{911} (\bibinfo{year}{2007}).

\bibitem[{\citenamefont{Oszl\'anyi et~al.}(1995)\citenamefont{Oszl\'anyi,
  Bortel, Faigel, Tegze, Gr\'an\'asy, Pekker, Stephens, Bendele, Dinnebier,
  Mih\'aly et~al.}}]{KC60_Dimer_PRB1995}
\bibinfo{author}{\bibfnamefont{G.}~\bibnamefont{Oszl\'anyi}},
  \bibinfo{author}{\bibfnamefont{G.}~\bibnamefont{Bortel}},
  \bibinfo{author}{\bibfnamefont{G.}~\bibnamefont{Faigel}},
  \bibinfo{author}{\bibfnamefont{M.}~\bibnamefont{Tegze}},
  \bibinfo{author}{\bibfnamefont{L.}~\bibnamefont{Gr\'an\'asy}},
  \bibinfo{author}{\bibfnamefont{S.}~\bibnamefont{Pekker}},
  \bibinfo{author}{\bibfnamefont{P.~W.} \bibnamefont{Stephens}},
  \bibinfo{author}{\bibfnamefont{G.}~\bibnamefont{Bendele}},
  \bibinfo{author}{\bibfnamefont{R.}~\bibnamefont{Dinnebier}},
  \bibinfo{author}{\bibfnamefont{G.}~\bibnamefont{Mih\'aly}},
  \bibnamefont{et~al.}, \bibinfo{journal}{Phys. Rev. B}
  \textbf{\bibinfo{volume}{51}}, \bibinfo{pages}{12228} (\bibinfo{year}{1995}).

\bibitem[{\citenamefont{Szirmai et~al.}(2011)\citenamefont{Szirmai,
  F\'{a}bi\'{a}n, D\'{o}ra, Koltai, Z\'{o}lyomi, K\"{u}rti, Nemes, Forr\'{o},
  and Simon}}]{SzirmaiPSSB2011}
\bibinfo{author}{\bibfnamefont{P.}~\bibnamefont{Szirmai}},
  \bibinfo{author}{\bibfnamefont{G.}~\bibnamefont{F\'{a}bi\'{a}n}},
  \bibinfo{author}{\bibfnamefont{B.}~\bibnamefont{D\'{o}ra}},
  \bibinfo{author}{\bibfnamefont{J.}~\bibnamefont{Koltai}},
  \bibinfo{author}{\bibfnamefont{V.}~\bibnamefont{Z\'{o}lyomi}},
  \bibinfo{author}{\bibfnamefont{J.}~\bibnamefont{K\"{u}rti}},
  \bibinfo{author}{\bibfnamefont{N.~M.} \bibnamefont{Nemes}},
  \bibinfo{author}{\bibfnamefont{L.}~\bibnamefont{Forr\'{o}}},
  \bibnamefont{and} \bibinfo{author}{\bibfnamefont{F.}~\bibnamefont{Simon}},
  \bibinfo{journal}{Phys. Status Solidi B} \textbf{\bibinfo{volume}{248}},
  \bibinfo{pages}{2688} (\bibinfo{year}{2011}).

\bibitem[{\citenamefont{Zhou et~al.}(1996)\citenamefont{Zhou, Watkins,
  McNamara~Rutledge, Messmer, and Chawla}}]{BDD_Zhou}
\bibinfo{author}{\bibfnamefont{X.}~\bibnamefont{Zhou}},
  \bibinfo{author}{\bibfnamefont{G.~D.} \bibnamefont{Watkins}},
  \bibinfo{author}{\bibfnamefont{K.~M.} \bibnamefont{McNamara~Rutledge}},
  \bibinfo{author}{\bibfnamefont{R.~P.} \bibnamefont{Messmer}},
  \bibnamefont{and} \bibinfo{author}{\bibfnamefont{S.}~\bibnamefont{Chawla}},
  \bibinfo{journal}{Phys. Rev. B} \textbf{\bibinfo{volume}{54}},
  \bibinfo{pages}{7881} (\bibinfo{year}{1996}).

\bibitem[{\citenamefont{{Mizuochi} et~al.}(2006)\citenamefont{{Mizuochi},
  {Watanabe}, {Okushi}, {Yamasaki}, {Niitsuma}, and
  {Sekiguchi}}}]{BDD_Mizuochi3}
\bibinfo{author}{\bibfnamefont{N.}~\bibnamefont{{Mizuochi}}},
  \bibinfo{author}{\bibfnamefont{H.}~\bibnamefont{{Watanabe}}},
  \bibinfo{author}{\bibfnamefont{H.}~\bibnamefont{{Okushi}}},
  \bibinfo{author}{\bibfnamefont{S.}~\bibnamefont{{Yamasaki}}},
  \bibinfo{author}{\bibfnamefont{J.}~\bibnamefont{{Niitsuma}}},
  \bibnamefont{and}
  \bibinfo{author}{\bibfnamefont{T.}~\bibnamefont{{Sekiguchi}}},
  \bibinfo{journal}{Appl. Phys. Lett.} \textbf{\bibinfo{volume}{88}},
  \bibinfo{eid}{091912} (\bibinfo{year}{2006}).

\bibitem[{\citenamefont{J\'anossy et~al.}(1993)\citenamefont{J\'anossy,
  Chauvet, Pekker, Cooper, and Forr\'o}}]{JanossyPRL1993}
\bibinfo{author}{\bibfnamefont{A.}~\bibnamefont{J\'anossy}},
  \bibinfo{author}{\bibfnamefont{O.}~\bibnamefont{Chauvet}},
  \bibinfo{author}{\bibfnamefont{S.}~\bibnamefont{Pekker}},
  \bibinfo{author}{\bibfnamefont{J.~R.} \bibnamefont{Cooper}},
  \bibnamefont{and} \bibinfo{author}{\bibfnamefont{L.}~\bibnamefont{Forr\'o}},
  \bibinfo{journal}{Phys. Rev. Lett.} \textbf{\bibinfo{volume}{71}},
  \bibinfo{pages}{1091} (\bibinfo{year}{1993}).

\bibitem[{\citenamefont{Feher and Kip}(1955)}]{FeherKip}
\bibinfo{author}{\bibfnamefont{G.}~\bibnamefont{Feher}} \bibnamefont{and}
  \bibinfo{author}{\bibfnamefont{A.~F.} \bibnamefont{Kip}},
  \bibinfo{journal}{Phys. Rev.} \textbf{\bibinfo{volume}{98}},
  \bibinfo{pages}{337} (\bibinfo{year}{1955}).

\bibitem[{\citenamefont{Mizuochi
  et~al.}(2004{\natexlab{a}})\citenamefont{Mizuochi, Ogura, Watanabe, Isoya,
  Okushi, and Yamasaki}}]{BDD_Mizuochi2}
\bibinfo{author}{\bibfnamefont{N.}~\bibnamefont{Mizuochi}},
  \bibinfo{author}{\bibfnamefont{M.}~\bibnamefont{Ogura}},
  \bibinfo{author}{\bibfnamefont{H.}~\bibnamefont{Watanabe}},
  \bibinfo{author}{\bibfnamefont{J.}~\bibnamefont{Isoya}},
  \bibinfo{author}{\bibfnamefont{H.}~\bibnamefont{Okushi}}, \bibnamefont{and}
  \bibinfo{author}{\bibfnamefont{S.}~\bibnamefont{Yamasaki}},
  \bibinfo{journal}{Diam. Relat. Mater.} \textbf{\bibinfo{volume}{13}},
  \bibinfo{pages}{2096 } (\bibinfo{year}{2004}{\natexlab{a}}).

\bibitem[{\citenamefont{Simon et~al.}({2001})\citenamefont{Simon, J\'{a}nossy,
  Feh\'{e}r, Mur\'{a}nyi, Garaj, Forr\'{o}, Petrovic, Bud'ko, Lapertot, Kogan
  et~al.}}]{SimonPRL2001}
\bibinfo{author}{\bibfnamefont{F.}~\bibnamefont{Simon}},
  \bibinfo{author}{\bibfnamefont{A.}~\bibnamefont{J\'{a}nossy}},
  \bibinfo{author}{\bibfnamefont{T.}~\bibnamefont{Feh\'{e}r}},
  \bibinfo{author}{\bibfnamefont{F.}~\bibnamefont{Mur\'{a}nyi}},
  \bibinfo{author}{\bibfnamefont{S.}~\bibnamefont{Garaj}},
  \bibinfo{author}{\bibfnamefont{L.}~\bibnamefont{Forr\'{o}}},
  \bibinfo{author}{\bibfnamefont{C.}~\bibnamefont{Petrovic}},
  \bibinfo{author}{\bibfnamefont{S.~L.} \bibnamefont{Bud'ko}},
  \bibinfo{author}{\bibfnamefont{G.}~\bibnamefont{Lapertot}},
  \bibinfo{author}{\bibfnamefont{V.~G.} \bibnamefont{Kogan}},
  \bibnamefont{et~al.}, \bibinfo{journal}{{Phys. Rev. Lett.}}
  \textbf{\bibinfo{volume}{{87}}} (\bibinfo{year}{{2001}}).

\bibitem[{\citenamefont{Simon et~al.}(2005)\citenamefont{Simon, J\'anossy,
  Feh\'er, Mur\'anyi, Garaj, Forr\'o, Petrovic, Bud'ko, Ribeiro, and
  Canfield}}]{SimonPRB2005b}
\bibinfo{author}{\bibfnamefont{F.}~\bibnamefont{Simon}},
  \bibinfo{author}{\bibfnamefont{A.}~\bibnamefont{J\'anossy}},
  \bibinfo{author}{\bibfnamefont{T.}~\bibnamefont{Feh\'er}},
  \bibinfo{author}{\bibfnamefont{F.}~\bibnamefont{Mur\'anyi}},
  \bibinfo{author}{\bibfnamefont{S.}~\bibnamefont{Garaj}},
  \bibinfo{author}{\bibfnamefont{L.}~\bibnamefont{Forr\'o}},
  \bibinfo{author}{\bibfnamefont{C.}~\bibnamefont{Petrovic}},
  \bibinfo{author}{\bibfnamefont{S.}~\bibnamefont{Bud'ko}},
  \bibinfo{author}{\bibfnamefont{R.~A.} \bibnamefont{Ribeiro}},
  \bibnamefont{and} \bibinfo{author}{\bibfnamefont{P.~C.}
  \bibnamefont{Canfield}}, \bibinfo{journal}{Phys. Rev. B}
  \textbf{\bibinfo{volume}{72}}, \bibinfo{pages}{012511}
  (\bibinfo{year}{2005}).

\bibitem[{\citenamefont{Gunnarsson}(1997)}]{GunnRMP}
\bibinfo{author}{\bibfnamefont{O.}~\bibnamefont{Gunnarsson}},
  \bibinfo{journal}{Rev. Mod. Phys.} \textbf{\bibinfo{volume}{69}},
  \bibinfo{pages}{575} (\bibinfo{year}{1997}).

\bibitem[{\citenamefont{Dresselhaus and Dresselhaus}(2002)}]{DresselhausAP2002}
\bibinfo{author}{\bibfnamefont{M.~S.} \bibnamefont{Dresselhaus}}
  \bibnamefont{and}
  \bibinfo{author}{\bibfnamefont{G.}~\bibnamefont{Dresselhaus}},
  \bibinfo{journal}{Adv. Phys.} \textbf{\bibinfo{volume}{51}},
  \bibinfo{pages}{1} (\bibinfo{year}{2002}).

\bibitem[{\citenamefont{F\'abi\'an et~al.}(2012)\citenamefont{F\'abi\'an,
  D\'ora, Antal, Szolnoki, Korecz, Rockenbauer, Nemes, Forr\'o, and
  Simon}}]{Fabian2012}
\bibinfo{author}{\bibfnamefont{G.}~\bibnamefont{F\'abi\'an}},
  \bibinfo{author}{\bibfnamefont{B.}~\bibnamefont{D\'ora}},
  \bibinfo{author}{\bibfnamefont{A.}~\bibnamefont{Antal}},
  \bibinfo{author}{\bibfnamefont{L.}~\bibnamefont{Szolnoki}},
  \bibinfo{author}{\bibfnamefont{L.}~\bibnamefont{Korecz}},
  \bibinfo{author}{\bibfnamefont{A.}~\bibnamefont{Rockenbauer}},
  \bibinfo{author}{\bibfnamefont{N.~M.} \bibnamefont{Nemes}},
  \bibinfo{author}{\bibfnamefont{L.}~\bibnamefont{Forr\'o}}, \bibnamefont{and}
  \bibinfo{author}{\bibfnamefont{F.}~\bibnamefont{Simon}},
  \bibinfo{journal}{Phys. Rev. B} \textbf{\bibinfo{volume}{85}},
  \bibinfo{pages}{235405} (\bibinfo{year}{2012}).

\bibitem[{\citenamefont{Mare\v{s} et~al.}(2006)\citenamefont{Mare\v{s},
  Hub\'{i}k, Nesl\'{a}dek, Kindl, and Kri\v{s}tofik}}]{BDD_Mares}
\bibinfo{author}{\bibfnamefont{J.}~\bibnamefont{Mare\v{s}}},
  \bibinfo{author}{\bibfnamefont{P.}~\bibnamefont{Hub\'{i}k}},
  \bibinfo{author}{\bibfnamefont{M.}~\bibnamefont{Nesl\'{a}dek}},
  \bibinfo{author}{\bibfnamefont{D.}~\bibnamefont{Kindl}}, \bibnamefont{and}
  \bibinfo{author}{\bibfnamefont{J.}~\bibnamefont{Kri\v{s}tofik}},
  \bibinfo{journal}{Diam. Relat. Mater.} \textbf{\bibinfo{volume}{15}},
  \bibinfo{pages}{1863 } (\bibinfo{year}{2006}).

\bibitem[{\citenamefont{Ortolani et~al.}(2006)\citenamefont{Ortolani, Lupi,
  Baldassarre, Schade, Calvani, Takano, Nagao, Takenouchi, and
  Kawarada}}]{BDD_Ortolani2006}
\bibinfo{author}{\bibfnamefont{M.}~\bibnamefont{Ortolani}},
  \bibinfo{author}{\bibfnamefont{S.}~\bibnamefont{Lupi}},
  \bibinfo{author}{\bibfnamefont{L.}~\bibnamefont{Baldassarre}},
  \bibinfo{author}{\bibfnamefont{U.}~\bibnamefont{Schade}},
  \bibinfo{author}{\bibfnamefont{P.}~\bibnamefont{Calvani}},
  \bibinfo{author}{\bibfnamefont{Y.}~\bibnamefont{Takano}},
  \bibinfo{author}{\bibfnamefont{M.}~\bibnamefont{Nagao}},
  \bibinfo{author}{\bibfnamefont{T.}~\bibnamefont{Takenouchi}},
  \bibnamefont{and} \bibinfo{author}{\bibfnamefont{H.}~\bibnamefont{Kawarada}},
  \bibinfo{journal}{Phys. Rev. Lett.} \textbf{\bibinfo{volume}{97}},
  \bibinfo{pages}{097002} (\bibinfo{year}{2006}).

\bibitem[{\citenamefont{Bourgeois et~al.}(2006)\citenamefont{Bourgeois,
  Bustarret, Achatz, Omn\`es, and Blase}}]{BDD_Bourgeois}
\bibinfo{author}{\bibfnamefont{E.}~\bibnamefont{Bourgeois}},
  \bibinfo{author}{\bibfnamefont{E.}~\bibnamefont{Bustarret}},
  \bibinfo{author}{\bibfnamefont{P.}~\bibnamefont{Achatz}},
  \bibinfo{author}{\bibfnamefont{F.}~\bibnamefont{Omn\`es}}, \bibnamefont{and}
  \bibinfo{author}{\bibfnamefont{X.}~\bibnamefont{Blase}},
  \bibinfo{journal}{Phys. Rev. B} \textbf{\bibinfo{volume}{74}},
  \bibinfo{pages}{094509} (\bibinfo{year}{2006}).

\bibitem[{\citenamefont{Monod and Beuneu}(1979)}]{MonodBeuneuPRB1979}
\bibinfo{author}{\bibfnamefont{P.}~\bibnamefont{Monod}} \bibnamefont{and}
  \bibinfo{author}{\bibfnamefont{F.}~\bibnamefont{Beuneu}},
  \bibinfo{journal}{Phys. Rev. B} \textbf{\bibinfo{volume}{19}},
  \bibinfo{pages}{911} (\bibinfo{year}{1979}).

\bibitem[{\citenamefont{Zhang et~al.}(2011)\citenamefont{Zhang, Janssens,
  Vanacken, Timmermans, Vac\'{\i}k, Ataklti, Decelle, Gillijns, Goderis, Haenen
  et~al.}}]{BDD_Zhang}
\bibinfo{author}{\bibfnamefont{G.}~\bibnamefont{Zhang}},
  \bibinfo{author}{\bibfnamefont{S.~D.} \bibnamefont{Janssens}},
  \bibinfo{author}{\bibfnamefont{J.}~\bibnamefont{Vanacken}},
  \bibinfo{author}{\bibfnamefont{M.}~\bibnamefont{Timmermans}},
  \bibinfo{author}{\bibfnamefont{J.}~\bibnamefont{Vac\'{\i}k}},
  \bibinfo{author}{\bibfnamefont{G.~W.} \bibnamefont{Ataklti}},
  \bibinfo{author}{\bibfnamefont{W.}~\bibnamefont{Decelle}},
  \bibinfo{author}{\bibfnamefont{W.}~\bibnamefont{Gillijns}},
  \bibinfo{author}{\bibfnamefont{B.}~\bibnamefont{Goderis}},
  \bibinfo{author}{\bibfnamefont{K.}~\bibnamefont{Haenen}},
  \bibnamefont{et~al.}, \bibinfo{journal}{Phys. Rev. B}
  \textbf{\bibinfo{volume}{84}}, \bibinfo{pages}{214517}
  (\bibinfo{year}{2011}).

\bibitem[{\citenamefont{Fabian and Das~Sarma}(1998)}]{FabianPRL1998}
\bibinfo{author}{\bibfnamefont{J.}~\bibnamefont{Fabian}} \bibnamefont{and}
  \bibinfo{author}{\bibfnamefont{S.}~\bibnamefont{Das~Sarma}},
  \bibinfo{journal}{Phys. Rev. Lett.} \textbf{\bibinfo{volume}{81}},
  \bibinfo{pages}{5624} (\bibinfo{year}{1998}), \bibinfo{note}{the "hot-spot"
  model explains the enhanced spin-relaxation of metals if the Fermi surface
  crosses the Brillouin zone boundaries.}

\bibitem[{\citenamefont{Mandal et~al.}(2010)\citenamefont{Mandal, Naud,
  Williams, Bustarret, Omn\`{e}s, Rodi\`{e}re, Meunier, Saminadayar, and
  B\"{a}uerle}}]{BDD_Mandal2010}
\bibinfo{author}{\bibfnamefont{S.}~\bibnamefont{Mandal}},
  \bibinfo{author}{\bibfnamefont{C.}~\bibnamefont{Naud}},
  \bibinfo{author}{\bibfnamefont{O.~A.} \bibnamefont{Williams}},
  \bibinfo{author}{\bibfnamefont{E.}~\bibnamefont{Bustarret}},
  \bibinfo{author}{\bibfnamefont{F.}~\bibnamefont{Omn\`{e}s}},
  \bibinfo{author}{\bibfnamefont{P.}~\bibnamefont{Rodi\`{e}re}},
  \bibinfo{author}{\bibfnamefont{T.}~\bibnamefont{Meunier}},
  \bibinfo{author}{\bibfnamefont{L.}~\bibnamefont{Saminadayar}},
  \bibnamefont{and}
  \bibinfo{author}{\bibfnamefont{C.}~\bibnamefont{B\"{a}uerle}},
  \bibinfo{journal}{Nanotechnology} \textbf{\bibinfo{volume}{21}},
  \bibinfo{pages}{195303} (\bibinfo{year}{2010}).

\bibitem[{\citenamefont{Bernard
  et~al.}(2004{\natexlab{a}})\citenamefont{Bernard, Deneuville, and
  Muret}}]{Bernard2004}
\bibinfo{author}{\bibfnamefont{M.}~\bibnamefont{Bernard}},
  \bibinfo{author}{\bibfnamefont{A.}~\bibnamefont{Deneuville}},
  \bibnamefont{and} \bibinfo{author}{\bibfnamefont{P.}~\bibnamefont{Muret}},
  \bibinfo{journal}{Diam. Relat. Mater.} \textbf{\bibinfo{volume}{13}},
  \bibinfo{pages}{282 } (\bibinfo{year}{2004}{\natexlab{a}}).

\bibitem[{\citenamefont{Bernard
  et~al.}(2004{\natexlab{b}})\citenamefont{Bernard, Baron, and
  Deneuville}}]{BDD_Bernard2004_DandRelMat}
\bibinfo{author}{\bibfnamefont{M.}~\bibnamefont{Bernard}},
  \bibinfo{author}{\bibfnamefont{C.}~\bibnamefont{Baron}}, \bibnamefont{and}
  \bibinfo{author}{\bibfnamefont{A.}~\bibnamefont{Deneuville}},
  \bibinfo{journal}{Diam. Relat. Mater.} \textbf{\bibinfo{volume}{13}},
  \bibinfo{pages}{896 } (\bibinfo{year}{2004}{\natexlab{b}}).

\bibitem[{\citenamefont{Sidorov and Ekimov}(2010)}]{BDD_Sidorov2010}
\bibinfo{author}{\bibfnamefont{V.~A.} \bibnamefont{Sidorov}} \bibnamefont{and}
  \bibinfo{author}{\bibfnamefont{E.~A.} \bibnamefont{Ekimov}},
  \bibinfo{journal}{Diam. Relat. Mater.} \textbf{\bibinfo{volume}{19}},
  \bibinfo{pages}{351 } (\bibinfo{year}{2010}).

\bibitem[{\citenamefont{Pruvost and Deneuville}(2001)}]{BDD_Pruvost}
\bibinfo{author}{\bibfnamefont{F.}~\bibnamefont{Pruvost}} \bibnamefont{and}
  \bibinfo{author}{\bibfnamefont{A.}~\bibnamefont{Deneuville}},
  \bibinfo{journal}{Diam. Relat. Mater.} \textbf{\bibinfo{volume}{10}},
  \bibinfo{pages}{531 } (\bibinfo{year}{2001}).

\bibitem[{\citenamefont{{Vlasov} et~al.}(2008)\citenamefont{{Vlasov}, {Ekimov},
  {Basov}, {Goovaerts}, and {Zoteev}}}]{BDD_Vlasov}
\bibinfo{author}{\bibfnamefont{I.~I.} \bibnamefont{{Vlasov}}},
  \bibinfo{author}{\bibfnamefont{E.~A.} \bibnamefont{{Ekimov}}},
  \bibinfo{author}{\bibfnamefont{A.~A.} \bibnamefont{{Basov}}},
  \bibinfo{author}{\bibfnamefont{E.}~\bibnamefont{{Goovaerts}}},
  \bibnamefont{and} \bibinfo{author}{\bibfnamefont{A.~V.}
  \bibnamefont{{Zoteev}}}, \bibinfo{journal}{ArXiv e-prints}
  (\bibinfo{year}{2008}), \eprint{0801.1611}.

\bibitem[{\citenamefont{Ferrari and Robertson}(2000)}]{FerrariPRB2000}
\bibinfo{author}{\bibfnamefont{A.~C.} \bibnamefont{Ferrari}} \bibnamefont{and}
  \bibinfo{author}{\bibfnamefont{J.}~\bibnamefont{Robertson}},
  \bibinfo{journal}{Phys. Rev. B} \textbf{\bibinfo{volume}{61}},
  \bibinfo{pages}{14095} (\bibinfo{year}{2000}).

\bibitem[{\citenamefont{Ferrari and Robertson}(2001)}]{FerrariPRB2001}
\bibinfo{author}{\bibfnamefont{A.~C.} \bibnamefont{Ferrari}} \bibnamefont{and}
  \bibinfo{author}{\bibfnamefont{J.}~\bibnamefont{Robertson}},
  \bibinfo{journal}{Phys. Rev. B} \textbf{\bibinfo{volume}{64}},
  \bibinfo{pages}{075414} (\bibinfo{year}{2001}).

\bibitem[{\citenamefont{Kresse and Furthm\"{u}ller}(1996)}]{KresseG_1996_2}
\bibinfo{author}{\bibfnamefont{G.}~\bibnamefont{Kresse}} \bibnamefont{and}
  \bibinfo{author}{\bibfnamefont{J.}~\bibnamefont{Furthm\"{u}ller}},
  \bibinfo{journal}{Phys. Rev. B} \textbf{\bibinfo{volume}{54}},
  \bibinfo{pages}{11169} (\bibinfo{year}{1996}).

\bibitem[{\citenamefont{Mizuochi
  et~al.}(2004{\natexlab{b}})\citenamefont{Mizuochi, Watanabe, Isoya, Okushi,
  and Yamasaki}}]{BDD_Mizuochi1}
\bibinfo{author}{\bibfnamefont{N.}~\bibnamefont{Mizuochi}},
  \bibinfo{author}{\bibfnamefont{H.}~\bibnamefont{Watanabe}},
  \bibinfo{author}{\bibfnamefont{J.}~\bibnamefont{Isoya}},
  \bibinfo{author}{\bibfnamefont{H.}~\bibnamefont{Okushi}}, \bibnamefont{and}
  \bibinfo{author}{\bibfnamefont{S.}~\bibnamefont{Yamasaki}},
  \bibinfo{journal}{Diam. Relat. Mater.} \textbf{\bibinfo{volume}{13}},
  \bibinfo{pages}{765 } (\bibinfo{year}{2004}{\natexlab{b}}).

\bibitem[{\citenamefont{Janssens et~al.}(2011)\citenamefont{Janssens,
  Pobedinskas, Vacik, Petr\'{a}kov\'{a}, Ruttens, D'Haen, Nesl\'{a}dek, Haenen,
  and Wagner}}]{BDD_Janssens}
\bibinfo{author}{\bibfnamefont{S.~D.} \bibnamefont{Janssens}},
  \bibinfo{author}{\bibfnamefont{P.}~\bibnamefont{Pobedinskas}},
  \bibinfo{author}{\bibfnamefont{J.}~\bibnamefont{Vacik}},
  \bibinfo{author}{\bibfnamefont{V.}~\bibnamefont{Petr\'{a}kov\'{a}}},
  \bibinfo{author}{\bibfnamefont{B.}~\bibnamefont{Ruttens}},
  \bibinfo{author}{\bibfnamefont{J.}~\bibnamefont{D'Haen}},
  \bibinfo{author}{\bibfnamefont{M.}~\bibnamefont{Nesl\'{a}dek}},
  \bibinfo{author}{\bibfnamefont{K.}~\bibnamefont{Haenen}}, \bibnamefont{and}
  \bibinfo{author}{\bibfnamefont{P.}~\bibnamefont{Wagner}},
  \bibinfo{journal}{New J. Phys.} \textbf{\bibinfo{volume}{13}},
  \bibinfo{pages}{083008} (\bibinfo{year}{2011}).

\bibitem[{\citenamefont{Liao et~al.}(1999)\citenamefont{Liao, Wang, and
  Yang}}]{BDD_Liao}
\bibinfo{author}{\bibfnamefont{C.}~\bibnamefont{Liao}},
  \bibinfo{author}{\bibfnamefont{Y.}~\bibnamefont{Wang}}, \bibnamefont{and}
  \bibinfo{author}{\bibfnamefont{S.}~\bibnamefont{Yang}},
  \bibinfo{journal}{Diam. Relat. Mater.} \textbf{\bibinfo{volume}{8}},
  \bibinfo{pages}{1229 } (\bibinfo{year}{1999}).

\bibitem[{Sig()}]{Sigma}
\bibinfo{note}{Note that this gives a lower bound for the penetration depth, as
  inter-grain interactions increase the resistivity.}

\bibitem[{\citenamefont{Abragam and Bleaney}(1970)}]{AbragamBleaneyBook}
\bibinfo{author}{\bibfnamefont{A.}~\bibnamefont{Abragam}} \bibnamefont{and}
  \bibinfo{author}{\bibfnamefont{B.}~\bibnamefont{Bleaney}},
  \emph{\bibinfo{title}{Electron paramagnetic resonance of transition ions}}
  (\bibinfo{publisher}{Oxford University Press}, \bibinfo{address}{Oxford,
  England}, \bibinfo{year}{1970}).

\end{thebibliography}

\end{document}